\documentclass[twocolumn]{aastex63}
\usepackage{amsmath}
\usepackage{here}

\newcommand{\gas}{\mathrm{g}}
\newcommand{\dst}{\mathrm{d}}

\newcommand{\sigmag}{\Sigma_{\gas}}

\newcommand{\sigmad}{\Sigma_{\dst}}

\newcommand{\tstop}{t_{\mathrm{stop}}}

\newcommand{\taus}{\tau_{\mathrm{s}}}

\newcommand{\cs}{c_{\mathrm{s}}}
\newcommand{\cd}{c_{\mathrm{d}}}
\newcommand{\hd}{H_{\mathrm{d}}}
\newcommand{\vk}{v_{\mathrm{K}}}

\newcommand{\vpp}{v_{\mathrm{pp}}}
\newcommand{\vfrag}{v_{\mathrm{frag}}}
\newcommand{\lmfp}{l_{\mathrm{mfp}}}
\newcommand{\mpar}{m_{\mathrm{p}}}
\newcommand{\rhoint}{\rho_{\mathrm{int}}}

\newcommand{\ddr}[1]{\frac{\partial #1}{\partial r}}
\newcommand{\ddt}[1]{\frac{\partial #1}{\partial t}}

\newcommand{\sigmagmmsn}{\Sigma_{\mathrm{g,MMSN}}}
\newcommand{\epmid}{\varepsilon_{\mathrm{mid}}}
\newcommand{\epmidwoBR}{\varepsilon_{\mathrm{mid,0}}}
\newcommand{\epsub}{\varepsilon_{\mathrm{s}}}

\newcommand{\alpeff}{\alpha_{\mathrm{eff}}}
\def\msun{M_{\odot}}
\def\merth{M_{\oplus}}

\received{--}
\accepted{--}
\submitjournal{ApJ}

\shorttitle{Effects of fragmentation and backreaction on coagulation instability}
\shortauthors{Tominaga et al.}

\begin{document}

\title{Nonlinear Outcome of Coagulation Instability in Protoplanetary Disks II: \\ Dust Ring Formation Mediated by Backreaction and Fragmentation}

\correspondingauthor{Ryosuke T. Tominaga}
\email{ryosuke.tominaga@riken.jp}
\author[0000-0002-8596-3505]{Ryosuke T. Tominaga}
\affiliation{RIKEN Cluster for Pioneering Research, 2-1 Hirosawa, Wako, Saitama 351-0198, Japan}

\author[0000-0001-9659-658X]{Hidekazu Tanaka}
\affiliation{Astronomical Institute, Graduate School of Science, Tohoku University, 6-3, Aramaki, Aoba-ku, Sendai 980-8578, Japan}

\author[0000-0001-8808-2132]{Hiroshi Kobayashi}
\affiliation{Department of Physics, Nagoya University, Nagoya, Aichi 464-8692, Japan}

\author[0000-0003-4366-6518]{Shu-ichiro Inutsuka}
\affiliation{Department of Physics, Nagoya University, Nagoya, Aichi 464-8692, Japan}



\begin{abstract}
In our previous work (Paper I), we demonstrated that coagulation instability results in dust concentration against depletion due to the radial drift and accelerates dust growth locally. In this work (Paper II), we perform numerical simulations of coagulation instability taking into account effects of backreaction to gas and collisional fragmentation of dust grains. We find that the slowdown of the dust drift due to backreaction regulates dust concentration in the nonlinear growth phase of coagulation instability. The dust-to-gas surface density ratio increases from $10^{-3}$ up to $\sim10^{-2}$. Each resulting dust ring tends to have mass of $\simeq0.5M_{\oplus}-1.5M_{\oplus}$ in our disk model. In contrast to Paper I, the dust surface density profile shows a local plateau structure at each dust ring. In spite of the regulation at the nonlinear growth, the efficient dust concentration reduces their collision velocity. As a result, dust grains can grow beyond the fragmentation barrier, and the dimensionless stopping time reaches unity as in Paper I. The necessary condition for the efficient dust growth is (1) weak turbulence of $\alpha<1\times10^{-3}$ and (2) a large critical velocity for dust fragmentation ($> 1$ m/s). The efficient dust concentration in outer regions will reduce the inward pebble flux and is expected to decelerate the planet formation via the pebble accretion. We also find that the resulting rings can be unstable to secular gravitational instability (GI). The subsequent secular GI promotes planetesimal formation. We thus expect that a combination of these instabilities is a promising mechanism for dust-ring and planetesimal formation.
\end{abstract}



\section{Introduction}\label{sec:intro}
Planetesimal formation from dust grains is the first step in the planet forming process. However, the origin of planetesimals is still under debate since there are processes known to inhibit planetesimal formation. The fast radial drift of dust grains due to aerodynamical dust-gas interaction is one issue \citep[e.g.,][]{Weidenschilling1977}. This radial drift causes depletion of dust and delays collisional dust growth. The resulting maximum dust size is limited up to meter size at a few au \citep[e.g., see][]{Brauer2008}. In order to avoid the drift barrier and enable planetesimal formation, two mechanisms have been proposed. One is porous dust aggregation \citep[e.g.,][]{Ormel2007a,Okuzumi2012,Kataoka2013b,Krijt2016,Arakawa2016,Garcia2020,Kobayashi2021} and the other is hydrodynamical clumping via streaming instability \citep[e.g.,][]{YoudinGoodman2005,Youdin2007,Johansen2007,Johansen2007nature,Krapp2019,Chen2020,Umurhan2020,Paardekooper2020,Paardekooper2021,McNally2021,Zhu2021,Yang2021,Carrera2021,Carrera2022} and other dust-gas instabilities including resonant-drag instability \citep[e.g.,][]{Squire2018a,Squire2018b,Zhuravlev2019,Zhuravlev2020} and secular gravitational instability \cite[e.g.,][]{Ward2000,Youdin2005a,Youdin2005b,Youdin2011,Shariff2011,Takeuchi_Ida2012,Michikoshi2012,Takahashi2014,Tominaga2018,Tominaga2019,Tominaga2020,Pierens2021}. Coagulation instability, proposed in \cite{Tominaga2021}, is one example of the latter mechanisms and is the main focus of this paper. In the companion paper \citep[][hereafter Paper I]{Tominaga2022}, we present the first numerical simulations and demonstrate that the instability operates and accelerates dust growth in dust-concentrated regions even after dust-to-gas ratio decreases down to $\sim10^{-3}$.

Another issue is dust fragmentation \citep[e.g.,][]{Weidenschilling1993,Blum2000,Blum2008,Guttler2010}, which is not treated in Paper I. According to $N$-body simulations of dust aggregate collisions \citep[e.g.,][]{Wada2009,Wada2013,Hasegawa2021}, a dust aggregate becomes larger via  collisions with collision velocities lower than a few to 10 m/s for silicate aggregates and 30-100 m/s for $\mathrm{H}_2\mathrm{O}$ ice if a monomer size is $0.1\;\mu\mathrm{m}$. If the monomer size is larger, the critical energy for separating monomers in contact becomes smaller \citep[e.g.,][]{Johnson1971,Chokshi1993}, which makes the critical velocity smaller \citep[e.g.,][]{Dominik1997,Wada2007,Wada2008}. According to a recent study by \citet{Tazaki2022}, optical and near-infrared polarimetric observations suggest that a monomer size is no greater than 0.4 $\mu\mathrm{m}$. Thus, the critical velocity can be roughly three times lower than the above value \citep[e.g., see Equation (8) in][]{Wada2013}. Since the maximum drift velocity can be tens m/s \citep[e.g., 54 m/s in the minimum mass solar nebula;][]{Hayashi1981}, collisional fragmentation can limit silicate dust growth or growth of dust aggregates with larger monomers. We however note that the sticking properties of silicate and $\mathrm{H}_2\mathrm{O}$ ice seem under debate \citep[e.g.,][]{Kimura2015,Kimura2020,Steinpilz2019,Musiolik2019}. 

Recent studies indicate that the sticking efficiency of $\mathrm{CO}_2$ ice is lower than $\mathrm{H}_2\mathrm{O}$ \citep{Musiolik2016a,Musiolik2016b}. This indicates that dust growth is potentially prevented even in the outer region beyond the $\mathrm{H}_2\mathrm{O}$ snow line if dust aggregates are covered by $\mathrm{CO}_2$ ice mantles \cite[e.g.,][]{Pinilla2017,Okuzumi2019}. Although the sticking efficiency depends on a complicated surface state of dust \citep[e.g., morphology and composition, see][]{Kouchi2021}, it is important to investigate how and to what extent such a low sticking efficiency affects the evolution of dust and disks.

To circumvent the fragmentation barrier, we need some processes to reduce the dust collision velocity. One possible process is backreaction in a dust-rich region. The dust drift velocity is reduced by the backreaction when the local dust-to-gas ratio becomes larger than unity \citep[e.g.,][]{Nakagawa1986}, and thus the collision velocity due to the differential drift is reduced. It is also possible that turbulence-induced collision velocity is reduced in a dust-rich region. The analytical study by \citet{Takeuchi2012} and the numerical simulations by \citet{Schreiber2018} show that the strength of turbulence driven by dust-gas instability, such as streaming instability \citep[e.g.,][]{YoudinGoodman2005}, decreases as the dust-to-gas ratio increases. Similar weakening of turbulence due to dust backreaction is also observed in hydrodynamics simulations of vertical shear instability (VSI) \citep[][]{Lin2019,Lehmann2021} and magnetohydrodynamics (MHD) simulations \citep{Xu2022}. The simulations show the enhancement of dust settling as the dust-to-gas surface density ratio increases, which may indicate that the turbulence-induced collision velocity is reduced. Thus, dust concentration is one promising pathway for dust grains to overcome the fragmentation barrier. Since the vertical concentration is limited by turbulence before the weakening (i.e., low dust-abundance limit), the key process is ``radial" dust concentration \citep[see also][]{Sekiya1998,Youdin2002}. Radial dust concentration at a pressure bump \citep[e.g.,][]{Whipple1972,Kato2012,Taki2016,Pinilla2021} or around the snow line \citep[e.g.,][]{Stevenson1988,Brauer2008b,Drazkowska2014,Schoonenberg2018} are examples of promising processes. The self-induced dust trap \citep{Gonzalez2017} is another process that promotes dust growth.

This work extends Paper I and discusses radial dust concentration via coagulation instability. We conduct numerical simulations and investigate to what extent dust grains grow and whether or not dust grains overcome the fragmentation barrier under the action of coagulation instability. We note that coagulation instability is distinct from the self-induced dust trap process since the backreaction is not prerequisite for the instability \citep[][see also Paper I]{Tominaga2021}. As explained below, we assume a steady global gas profile while we treat reduction of dust drift, collision velocities, and turbulence strength due to the backreaction. This steady-gas assumption is validated since coagulation instability is essentially a one-fluid instability as shown in \citet{Tominaga2021}. By performing such simplified simulations, we can focus on the effects of fragmentation and backreaction on coagulation instability.

 This paper is organized as follows. We describe basic equations, numerical methods, and disk models in Section \ref{sec:method}. In Section \ref{sec:results}, we first present results of a fiducial run (Section \ref{subsec:fiducial}) and next show results of parameter studies (Section \ref{subsec:params}). In Section \ref{sec:discussion}, we discuss relevant critical fragmentation velocity and turbulence strength indicated in the previous studies, a possible combined process with the self-induced dust trap \citep[][]{Gonzalez2017}, effects on pebble accretion, and the development to other disk instabilities, especially to secular gravitational instability \citep[e.g.,][]{Ward2000,Youdin2011,Takahashi2014,Tominaga2019}. We give a summary in Section \ref{sec:summary}

\section{Methods and Models}\label{sec:method}
As in Paper I, we perform one-dimensional simulations and discuss dust evolution in a steady axisymmetric gas disk around a one-solar-mass star. We solve the following moment equations formulated by \citet{Sato2016}:
\begin{equation}
\ddt{\sigmad}+\frac{1}{r}\ddr{ }\left(r\sigmad v_r\right)=0,\label{eq:dusteoc}
\end{equation}
\begin{equation}
v_r\equiv\left<v_r\right>-\frac{D}{\sigmad}\ddr{\sigmad},\label{eq:vr}
\end{equation}
\begin{equation}
\ddt{\mpar}+v_r\ddr{\mpar}=p_{\mathrm{eff}}\frac{2\sqrt{\pi}a^2\Delta\vpp}{\hd}\sigmad,\label{eq:dmpdt}
\end{equation}
where we adopt the cylindrical coordinate $(r,\phi,z)$, $\sigmad$ is the dust surface density, $v_r$ is the advection velocity of dust, $\left<v_r\right>$ is the mean drift velocity, $D$ is the diffusion coefficient, $\mpar=4\pi\rhoint a^3/3$ is a single dust mass with an internal density $\rhoint$, $p_{\mathrm{eff}}$ is the sticking efficiency, $\Delta\vpp$ is the collision velocity, and $\hd$ is the dust scale height. The introduced dust mass $\mpar$ is the so-called peak-mass, which is one moment value of a dust size distribution and represents a mass-dominating dust size \citep[e.g.,][]{Estrada2008,Ormel2008,Sato2016}. As in the previous studies \citep[e.g.,][]{Sato2016,Taki2021} and Paper I, we consider the Brownian motion, differential drift velocities, and turbulence-induced collision velocity to calculate $\Delta \vpp$ (see Paper I and the reference therein). The adopted equations are almost the same as those in Paper I except for (A) $p_{\mathrm{eff}}$ on the right hand side of Equation (\ref{eq:dmpdt}), (B) a model of turbulence strength, and (C) a formula of the mean drift velocity $\left<v_r\right>$, which are explained below. 

\subsection{Sticking efficiency}
To take the effect of fragmentation into account, we follow \citet{OH2012} and \citet{Okuzumi2016} and adopt their model of the sticking efficiency $p_{\mathrm{eff}}$:
\begin{equation}
p_{\mathrm{eff}}\equiv \mathrm{min}\left(1, -\frac{\ln\left(\Delta\vpp/\vfrag\right)}{\ln 5}\right),\label{eq:peff}
\end{equation}
where $\vfrag$ is the critical fragmentation velocity. In this model, collisional fragmentation decreases the growth efficiency for $\Delta\vpp>\vfrag/5$. Collisions at a larger velocity than $\vfrag$ lead to catastrophic fragmentation, and net growth does not occur, i.e., $p_{\mathrm{eff}}\leq0$. The above formula comes from a fit to the results of numerical simulations of similar-sized dust aggregate collisions in \citet{Wada2009} \citep[see Figure 5 in][]{Okuzumi2016}. The critical fragmentation velocity depends on compositions (surface energy),  monomer sizes, and the mass ratio of colliding aggregates \citep[e.g.,][]{Chokshi1993,Hasegawa2021}. In this work, we simply treat $\vfrag$ as a parameter and assume it to be radially constant rather than adopting spatially varying dust compositions and monomer sizes. Although there must be multiple snow lines in a disk, we assume the constant $\vfrag$ in this first study since this simplification provides clear understanding of the effect of fragmentation on coagulation instability.

\subsection{Turbulence strength}
Some physical values are related to turbulence strength $\alpha$ \citep{Shakura1973}. For example, the dust diffusivity and the dust scale height are \citep[][]{YL2007}
\begin{equation}
D=\frac{1+\taus+4\taus^2}{(1+\taus^2)^2}\alpha\cs H,\label{eq:diffcoef}
\end{equation}
\begin{equation}
\hd=H\left(1+\frac{\taus}{\alpha}\left(\frac{1+2\taus}{1+\taus}\right)\right)^{-1/2},
\end{equation}
where $\cs$ is the sound speed, $H$ is the gas scale height, and $\taus\equiv\tstop\Omega$ represents the stopping time $\tstop$ normalized by the Keplerian angular velocity $\Omega=\sqrt{G\msun/r^3}$. We also assume isotropic turbulence. The turbulence-driven collision velocity $\Delta v_{\mathrm{t}}$ included in $\Delta\vpp$ also depends on the turbulence strength as $\Delta v_{\mathrm{t}}\propto\sqrt{\alpha}\cs$ \citep[see Equations (17) and (18) in][]{Ormel2007}.

As noted in Section \ref{sec:intro}, the previous studies have found that the turbulence strength was reduced by the backreaction \citep[e.g.,][]{Takeuchi2012,Schreiber2018,Xu2022}. As indicated in \citet{Takeuchi2012}, this leads to a feedback process once any processes such as coagulation instability increase the midplane dust-to-gas ratio to $\sim 1$ as follows. The backreaction becomes effective for the midplane dust-to-gas ratio of $\sim 1$ or larger, which decreases the turbulence strength $\alpha$. The smaller $\alpha$ enhances the dust vertical settling because the vertical diffusion becomes less efficient. This leads to a further increase in the midplane dust-to-gas ratio.

The efficiency of the above feedback process is determined by the dependence of $\alpha$ on the midplane dust-to-gas ratio. \cite{Takeuchi2012} showed analytically that $\alpha$ is roughly inversely proportional to the midplane dust-to-gas ratio $\epmid$ for $\epmid\gg1$ (see Equations (28) and (29) and Figures 2 and 3 therein). The local simulations by \citet{Schreiber2018} also show a similar trend. These studies considered only dust-gas instabilities as a source of turbulence. \citet{Lin2019} found that the backreaction reduces the efficiency of VSI and enhances dust settling \citep[see also][]{Lehmann2021}. \cite{Xu2022} found an enhancement of dust settling due to the backreaction to gas in the presence of MHD turbulence \citep[see also][]{Yang2018}. Thus, we can expect the weakening of hydrodynamic and MHD turbulence due to the backreaction as well as in the case of turbulence driven by dust-gas instabilities. Although the $\epmid$-dependence of $\alpha$ might be different for a different source of turbulence, we assume the $\epmid$-dependence found in \citet{Takeuchi2012} and \citet{Schreiber2018} for large $\epmid$ in this first study. To do so, we introduce an effective alpha $\alpha=\alpha_{\mathrm{eff}}$ defined as follows:
\begin{equation}
\alpha_{\mathrm{eff}}=\frac{\alpha_0}{1+\epmid},\label{eq:alpeff}
\end{equation}
where $\alpha_0$ is a parameter. This simple model of $\alpha_{\mathrm{eff}}$ is adopted in previous studies \citep[][]{Hyodo2019,Ida2021,Li2021b}. The parameter $\alpha_0$ mimics the possible existence of turbulence driven by some processes that do not require dust grains. Thus, the turbulence strength is given by $\alpha=\alpha_0$ in the limit of low dust-to-gas ratio or when we switch off the backreaction to the turbulence strength.

The turbulence strength decreases as the local dust-gas ratio in the dust layer increases, and the backreaction becomes effective. The timing at which the increase of the midplane dust-to-gas ratio accelerates is roughly the timing at which the midplane dust-to-gas ratio becomes $\sim 1$ in the absence of the backreaction (i.e., for $\alpha=\alpha_0$). In the following, we derive a simple formula of the midplane dust-to-gas ratio for $\alpha=\alpha_{\mathrm{eff}}$. Assuming the Gaussian profiles for both dust density $\rho_{\dst}$ and gas density $\rho_{\gas}$ (see Paper I), we relate $\epmid$ with the dust-to-gas surface density ratio $\varepsilon$:
\begin{equation}
\epmid = \frac{H}{\hd}\varepsilon, 
\end{equation}
\begin{equation}
\varepsilon\equiv\frac{\sigmad}{\sigmag},
\end{equation}
where the gas surface density is denoted by $\sigmag$. In the absence of the backreaction, the midplane dust-to-gas ratio, i.e. $\epmidwoBR$, is given by
\begin{equation}
\epmidwoBR=\sqrt{\frac{\taus}{\alpha_0}\frac{1+2\taus}{1+\taus}}\varepsilon,
\end{equation}
where we assume $\hd (\alpha_0)/H\simeq\sqrt{\alpha_0(1+\taus)/\taus(1+2\taus)}$. Assuming $\hd/H\simeq\sqrt{\alpha_{\mathrm{eff}}(1+\taus)/\taus(1+2\taus)}$ and using Equation (\ref{eq:alpeff}), one obtains
\begin{align}
\epmid &= \varepsilon\sqrt{(1+\epmid)\frac{\taus}{\alpha_0}\frac{1+2\taus}{1+\taus}},\notag\\
&=\epmidwoBR\sqrt{1+\epmid},
\end{align}
where $\epmidwoBR\equiv\varepsilon\sqrt{\taus/\alpha_0}$ denotes the midplane dust-to-gas ratio in the absence of the backreaction to the turbulence strength via the dust enrichment. We then arrive at the following equation for $\epmid$:
\begin{equation}
\epmid = \frac{\epmidwoBR^2}{2}\left(1+\sqrt{1+\frac{4}{\epmidwoBR^2}}\right). \label{eq:epmid_depend}
\end{equation}
One has $\epmid=\epmidwoBR$ for $\epmidwoBR\ll 1$, e.g., before the significant dust enrichment via coagulation instability. The increase of the midplane dust-to-gas ratio $\epmid$ accelerates once the dust enrichment makes the surface density ratio $\varepsilon$ so large that $\epmidwoBR$ becomes unity or larger \citep[see also Figure 3 in][]{Takeuchi2012}.

\begin{figure}[tp]
	\begin{center}
	\hspace{100pt}\raisebox{20pt}{
	\includegraphics[width=0.9\columnwidth]{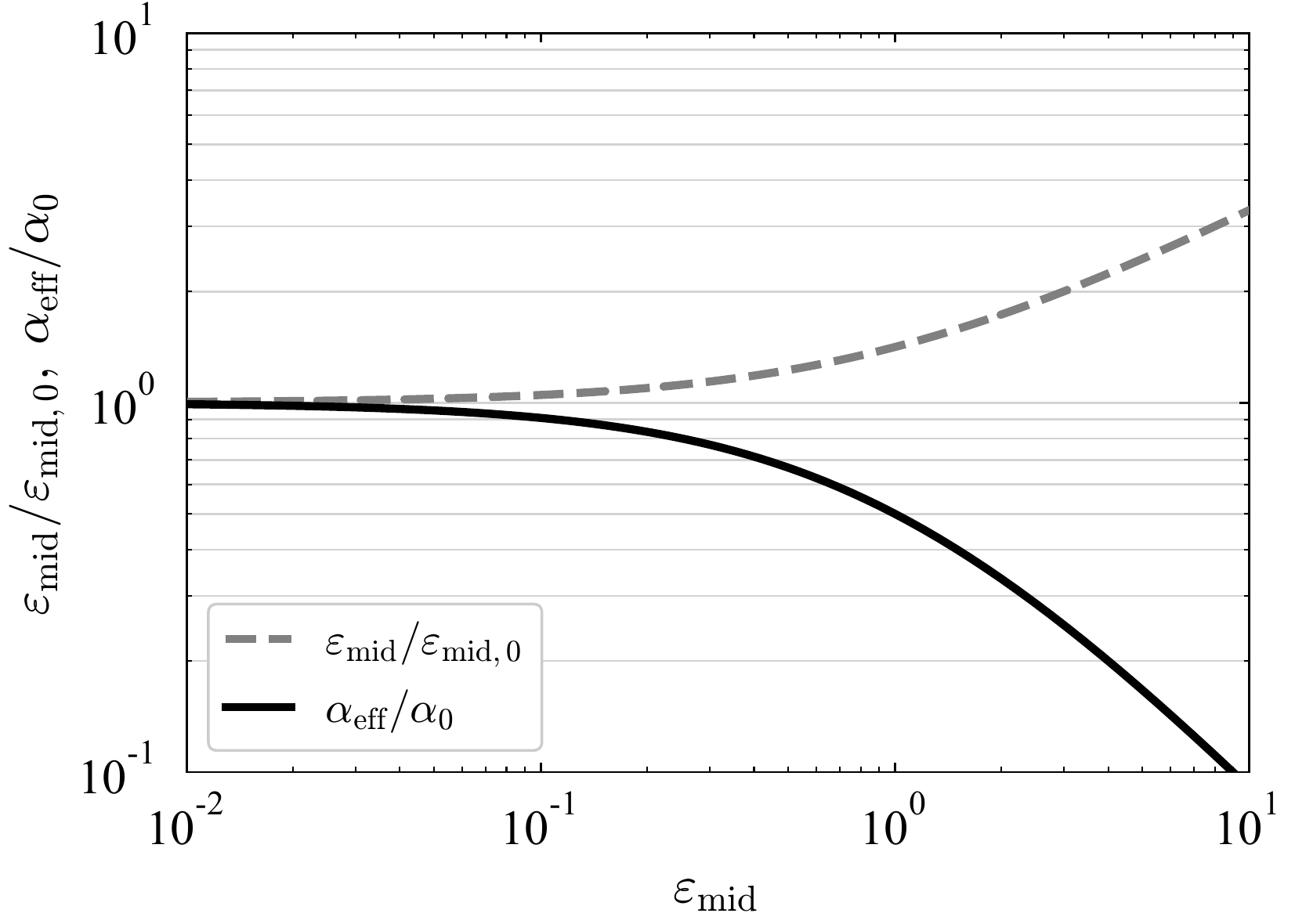} 
	}
	\end{center}
	\vspace{-30pt}
\caption{The ratios $\alpha_{\mathrm{eff}}/\alpha_0$ and $\epmid/\epmidwoBR$ as a function $\epmid$ (see also Equations (\ref{eq:alpeff}) and (\ref{eq:epmid_depend})). In most of the present simulations in Section \ref{sec:results}, the turbulence strength $\alpha_{\mathrm{eff}}$ becomes smaller than $\alpha_0$ by a factor of $\simeq2-5$ (see Figure \ref{fig:evol_fid}). From this figure, we see that $\epmid$ is in the range of $\simeq1-4$, and the difference between $\epmid$ and $\epmidwoBR$ is a few in such a case. } 
\label{fig:epmid_dependence}
\end{figure}

Figure \ref{fig:epmid_dependence} shows the ratios $\alpha_{\mathrm{eff}}/\alpha_0$ and $\epmid/\epmidwoBR$ as a function of $\epmid$. In the absence of the backreaction, the turbulence strength is independent of $\epmid$, i.e., $\alpha=\alpha_0$. The midplane dust-to-gas ratio $\epmidwoBR$ linearly increases as the surface density ratio $\varepsilon$ increases. In the presence of the backreaction, the increase of the midplane dust-to-gas ratio accelerates as $\varepsilon$ increases. The gray dashed line in Figure \ref{fig:epmid_dependence} compares $\epmid$ with $\epmidwoBR$. This line thus shows the impact of the backreaction on the dust settling. The difference between $\epmid$ and $\epmidwoBR$ is a factor of $\simeq3$ for $\epmid\simeq10$ (i.e., $\epmidwoBR\simeq3$). As the black line shows, the turbulence strength $\alpha_{\mathrm{eff}}$ becomes smaller than $\alpha_0$ by a factor of 10 once any processes increase $\varepsilon$ so that $\epmidwoBR$ becomes $\simeq3$.  As described in Section \ref{sec:results}, we observe a decrease in $\alpha_{\mathrm{eff}}$ by a factor of $\simeq 2-5$ in most runs. From Figure \ref{fig:epmid_dependence}, we can see that the backreaction causes a factor of a few times stronger dust settling in such cases. 

\subsection{Drift velocity in the dust layer}
We assume sub-Keplerian gas orbital velocity of $(1-\eta)\vk$, where $\vk=r\Omega=\sqrt{G\msun/r}$ is the Keplerian velocity. Because of the backreaction to the gas disk, the dust drift speed becomes smaller in a dust-rich region than in a dust-poor region \citep[e.g.,][]{Nakagawa1981}. This reduction of the drift velocity occurs in the dust sublayer around the midplane. To take this effect into account, we adopt the following mean drift velocity:
\begin{equation}
\left<v_r\right> \equiv -\frac{2\taus}{(1+\epsub)^2+\taus^2}\eta\vk,\label{eq:vrdrift}
\end{equation}
where $\epsub$ is a dust-to-gas ratio averaged in the dust sublayer. We assume the vertical extent of the sublayer to be $-\hd\leq z \leq \hd$ (see also Appendix \ref{app:zsdepend}). The ratio $\epsub$ takes a similar value to $\epmid$ once dust grains settle enough. In the present model, the dust-to-gas ratio in the dust sublayer is analytically given as follows:
\begin{align}
\epsub&\equiv\varepsilon\;\mathrm{erf}\left(\frac{1}{\sqrt{2}}\right)\left[\mathrm{erf}\left(\frac{\hd}{\sqrt{2}H}\right)\right]^{-1},\label{eq:epsub}\\
&=\epmid\frac{\hd}{H}\mathrm{erf}\left(\frac{1}{\sqrt{2}}\right)\left[\mathrm{erf}\left(\frac{\hd}{\sqrt{2}H}\right)\right]^{-1}
\end{align}

We calculate the other physical values in Equation (\ref{eq:vrdrift}) in the conventional way. The parameter $\eta$ is given by
\begin{equation}
\eta\equiv-\frac{1}{2\rho_{\gas}r\Omega^2}\frac{\partial\cs^2\rho_{\gas}}{\partial r}.\label{eq:eta}
\end{equation}
The sound speed of the temperature $T$ is given by $\cs=\sqrt{k_{\mathrm{B}}T/\mu m_{\mathrm{H}}}$, where $k_{\mathrm{B}}$ and $m_{\mathrm{H}}$ are the Boltzmann constant and the hydrogen mass, and $\mu=2.34$ is the mean molecular weight. We calculate $\eta$ using the midplane gas density and temperature (see Section \ref{subsec:diskmodel}). We use the Epstein law and the Stokes law to calculate the stopping time:
\begin{equation}
\tstop= 
\begin{cases}
\displaystyle\sqrt{\frac{\pi}{8}}\frac{\rhoint a}{\rho_{\gas}\cs}, &\displaystyle \left(\frac{a}{\lmfp}\leq\frac{9}{4}\right)\\
\displaystyle\sqrt{\frac{\pi}{8}}\frac{4\rhoint a^2}{9\rho_{\gas}\cs\lmfp}, &\displaystyle \left(\frac{a}{\lmfp}>\frac{9}{4}\right)
\end{cases}
\label{eq:def_taus}
\end{equation}
where $\rhoint=1.4\;\mathrm{g}\:\mathrm{cm}^{-3}$ is an internal mass density of dust grains, $\lmfp$ is the mean free path of gas. As for $\eta$, we use the midplane value of the gas density and the temperature to calculate $\tstop$ since most of dust grains reside around the midplane when coagulation instability becomes operational ($\taus\sim 0.1-1$).

The collision velocities due to the radial and azimuthal differential drift, $\Delta v_r,\;\Delta v_{\phi}$, are also reduced as $\epsub$ increases:
\begin{equation}
\Delta v_r = \left|-\frac{2\tau_{\mathrm{s},1}}{(1+\epsub)^2+\tau_{\mathrm{s},1}^2}+\frac{2\tau_{\mathrm{s},2}}{(1+\epsub)^2+\tau_{\mathrm{s},2}^2}\right|\eta\vk,\label{eq:vrcol}
\end{equation} 
\begin{equation}
\Delta v_{\phi}=\left|-\frac{1+\epsub}{(1+\epsub)^2+\tau_{\mathrm{s},1}^2}+\frac{1+\epsub}{(1+\epsub)^2+\tau_{\mathrm{s},2}^2}\right| \eta\vk,\label{eq:vphicol}
\end{equation}
where we consider a collision between dust grains of $\taus=\tau_{\mathrm{s},1}$ and $\tau_{\mathrm{s},2}$, and also use the following formula of the azimuthal velocity \citep[see][]{Nakagawa1986}:
\begin{equation}
v_{\phi} = \vk - \frac{1+\epsub}{(1+\epsub)^2+\taus^2}\eta\vk.
\end{equation}
Following the formalism by \citet{Sato2016}, we consider collisions of $\tau_{\mathrm{s},2}=0.5\tau_{\mathrm{s},1}$ throughout this paper as in Paper I. We refer readers to \citet{Sato2016} for the validation of this assumption as well as the derivation of the moment equations.

\subsection{Disk models}\label{subsec:diskmodel}

\begin{deluxetable}{c|cccccc}[ht]
\tablecaption{Summary of parameters}\label{tab:param}
\tablehead{
\colhead{Runs}  & \colhead{$\alpha_0$} & \colhead{$R_0$ [au]} & \colhead{$\Delta \vfrag$ [m/s]} 
}
\startdata
a10vf30BR & $1\times 10^{-3}$ & 10 & 30  \\
a5vf30BR & $5\times 10^{-4}$ & 50  & 30   \\
a3vf30BR & $3\times 10^{-4}$ & 50  & 30   \\
a1vf30BR & $1\times 10^{-4}$ & 50  & 30   \\
a05vf30BR & $5\times 10^{-5}$ & 50  & 30   \\
a03vf30BR & $3\times 10^{-5}$ & 50  & 30   \\ \hline
a5vf10BR & $5\times 10^{-4}$ & 10  & 10   \\
a3vf10BR & $3\times 10^{-4}$ & 50  & 10   \\
a1vf10BR & $1\times 10^{-4}$ & 50  & 10   \\
a05vf10BR & $5\times 10^{-5}$ & 50  & 10   \\
a03vf10BR & $3\times 10^{-5}$ & 50  & 10   \\ \hline
a3vf3BR & $3\times 10^{-4}$ & 10  & 3   \\
a1vf3BR & $1\times 10^{-4}$ & 10  & 3   \\
a05vf3BR & $5\times 10^{-5}$ & 50  & 3   \\
a03vf3BR & $3\times 10^{-5}$ & 50  & 3   \\ \hline
a1vf1BR & $1\times 10^{-4}$ & 10  & 1  \\
a05vf1BR & $5\times 10^{-5}$ & 10  & 1  \\
a03vf1BR & $3\times 10^{-5}$ & 50  & 1  \\
\enddata
\end{deluxetable}

We assume a steady gas surface density profile and a temperature profile as follows
\begin{equation}
\sigmag(r)=\Sigma_{\gas,10}\left(\frac{r}{10\;\mathrm{au}}\right)^{-1},
\end{equation}
\begin{equation}
T(r) = T_{100}\left(\frac{r}{100\;\mathrm{au}}\right)^{-1/2},
\end{equation}
where $\Sigma_{\gas,10}$ and $T_{100}$ are constants. The dependence of resulting ring locations on the power law index of $\sigmag$ is investigated in Paper I, where we adopt $d\ln \sigmag/d\ln r=-0.5, -1$, and $-1.5$. In Paper II, we adopt $d\ln \sigmag/d\ln r=-1$ as a fiducial case. This value is included in the inferred range for the observed disks \citep[][]{Kitamura2002,Andrews2009}. In this paper, we fix the constants $T_{100}$ and $\Sigma_{\gas,10}$ as $T_{100}=20\;\mathrm{K}$ and $\Sigma_{\gas,10}=2\sigmagmmsn(10\;\mathrm{au})$, where $\sigmagmmsn(10\;\mathrm{au})=53.75\;\mathrm{g\;cm}^{-2}$ is the gas surface density at $r=10\;\mathrm{au}$ in the minimum mass solar nebula disk \citep[MMSN;][]{Hayashi1981}. The initial inner and outer boundaries are located at $r=5\;\mathrm{au}$ and $100\;\mathrm{au}$, respectively. The gas disk mass is thus $\sim0.07\msun$, which is relatively massive among the observationally inferred disk masses \citep[e.g.,][]{Andrews2010,Barenfeld2016,Manara2018,Tychoniec2020,Mulders2021}. As discussed in Section \ref{subsec:secular}, massive disks are necessary for the onset of secular GI \citep[see also][]{Takahashi2014,Latter2017,Tominaga2020}. We find that simulations with the above disk mass show both (1) successful cases where the resulting rings becomes unstable to secular GI and (2) unsuccessful cases where the rings are stable to secular GI even after coagulation instability develops. We thus adopt this disk model.

We assume initial dust-to-gas ratio of $10^{-2}$, and thus the initial dust surface density profile is
\begin{equation}
\sigmad(r)=10^{-2}\Sigma_{\gas,10}\left(\frac{r}{10\;\mathrm{au}}\right)^{-1}.
\end{equation}
The dust disk mass is $\simeq2.4\times10^2\merth$, which is also relatively large among the observationally inferred masses \citep[e.g.,][]{Mulders2021}. We set initial dust size to be $10\;\mu\mathrm{m}$ as in Paper I in order to shorten duration of the initial size growth phase where dust grains hardly drift and coagulation instability does not develop.

We set initial perturbations almost in the same way as in Paper I: we displace dust cells using random perturbations at $t=0\;\mathrm{yr}$. We introduce a radius $R_0$ and input perturbations at $r\geq R_0$. In most runs where coagulation instability easily grows, we adopt $R_0=50\;\mathrm{au}$ to see the propagation of perturbations as in Paper I. If we use smaller $R_0$ in such cases, the nonlinear growth of coagulation instability quickly develops at inner radii, and $\taus=1$ is achieved well before outer dust grains start drifting. On the other hand, we adopt $R_0=10\;\mathrm{au}$ for simulations with small $\vfrag$ or large $\alpha$. These correspond to cases where coagulation instability grows less efficiently or is stabilized. In contrast to Paper I, we also investigate cases of low $\alpha_0$ ($\sim10^{-5}$) where shorter-wavelength perturbations will grow. We thus adopt 384 modes for initial perturbations to introduce shorter-wavelength perturbations than in Paper I that assumes 128 modes (see Section 4.1 in Paper I).

Labels of runs and parameters are listed in Table \ref{tab:param}. To highlight the fact that the effects of backreaction is included in contrast to runs in Paper I, we add ``BR" at the end of each label.

\subsection{Numerics}
The numerical method we adopt is the same as in Paper I. We use the Lagrangian-cell method to avoid numerical diffusion due to advection \citep{Tominaga2018}. We adopt the operator-splitting method \citep[e.g.,][]{Inoue2008} for time integration with the radial drift and diffusion parts, and use the second-order Runge-Kutta integrator. We utilize the super-time-stepping scheme to accelerate the time integration \citep[][]{Alexiades1996,Meyer2012,Meyer2014} once the radial diffusion limits the time step. We refer readers to Paper I for more detailed description.

As in Paper I, we stop simulations once $\taus=1$ is achieved in one dust concentrated region. Time evolution after $\taus=1$ leads to cell-crossing that our one-dimensional numerical method can not describe. Besides, the moment description adopted in this work might be inappropriate in such a case since the local dust size distribution will be bimodal. We calculate dust evolution for $2\times10^5\;\mathrm{yr}$ unless $\taus$ reaches unity.

\begin{figure}[tp]
	\begin{center}
	\hspace{100pt}\raisebox{20pt}{
	\includegraphics[width=0.9\columnwidth]{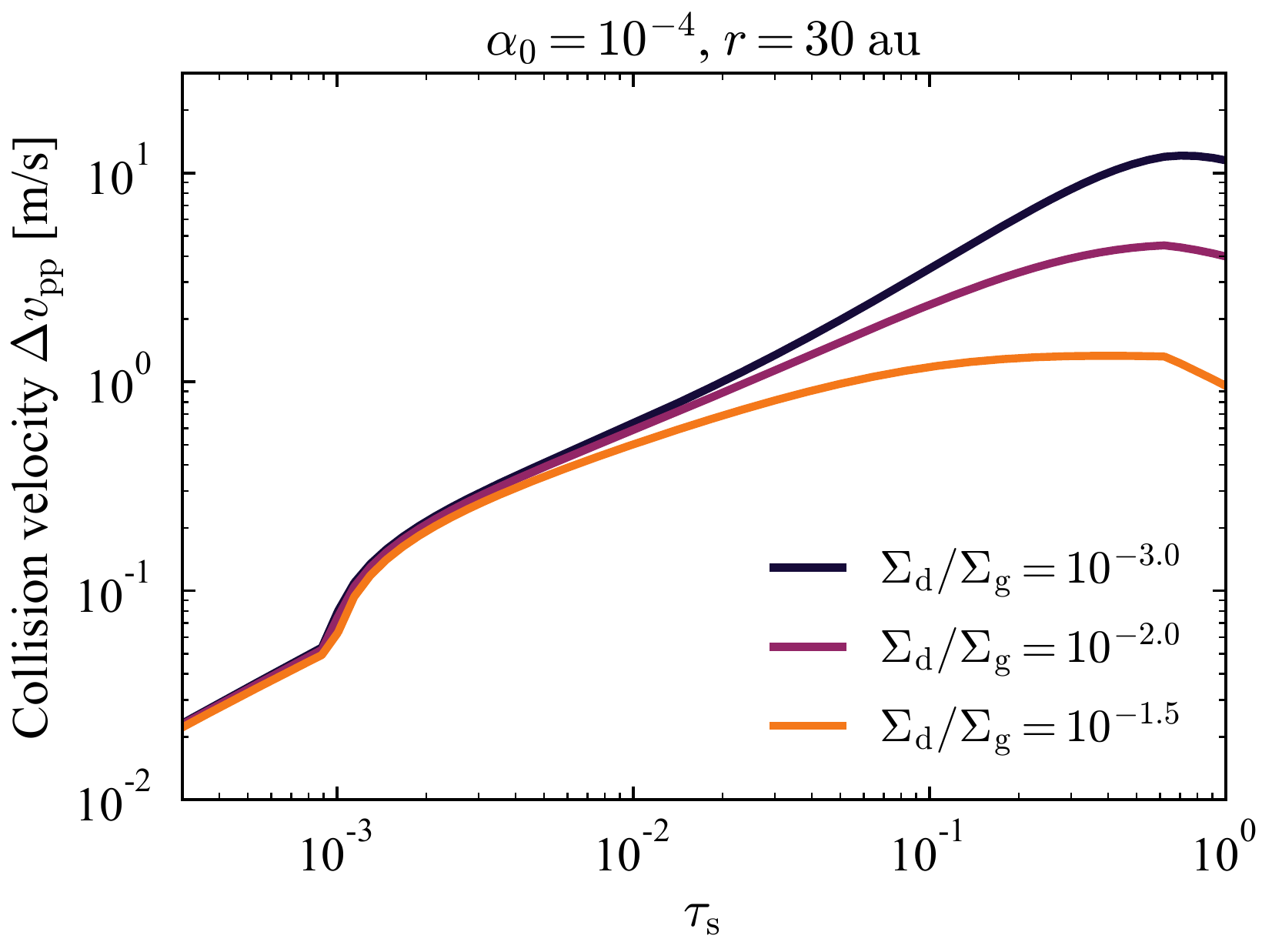} 
	}
	\end{center}
	\vspace{-30pt}
\caption{The collision velocity $\Delta\vpp$ as a function of $\taus$ for $\sigmad/\sigmag=10^{-3}$ (black line), $10^{-2}$ (purple line) and $10^{-1.5}$ (orange line). We assume $\alpha_0=1\times 10^{-4}$ for each case. }
\label{fig:ep_vcol_map}
\end{figure}

\begin{figure*}[tp]
	\begin{center}
	\hspace{0pt}\raisebox{20pt}{
	\includegraphics[width=1.5\columnwidth]{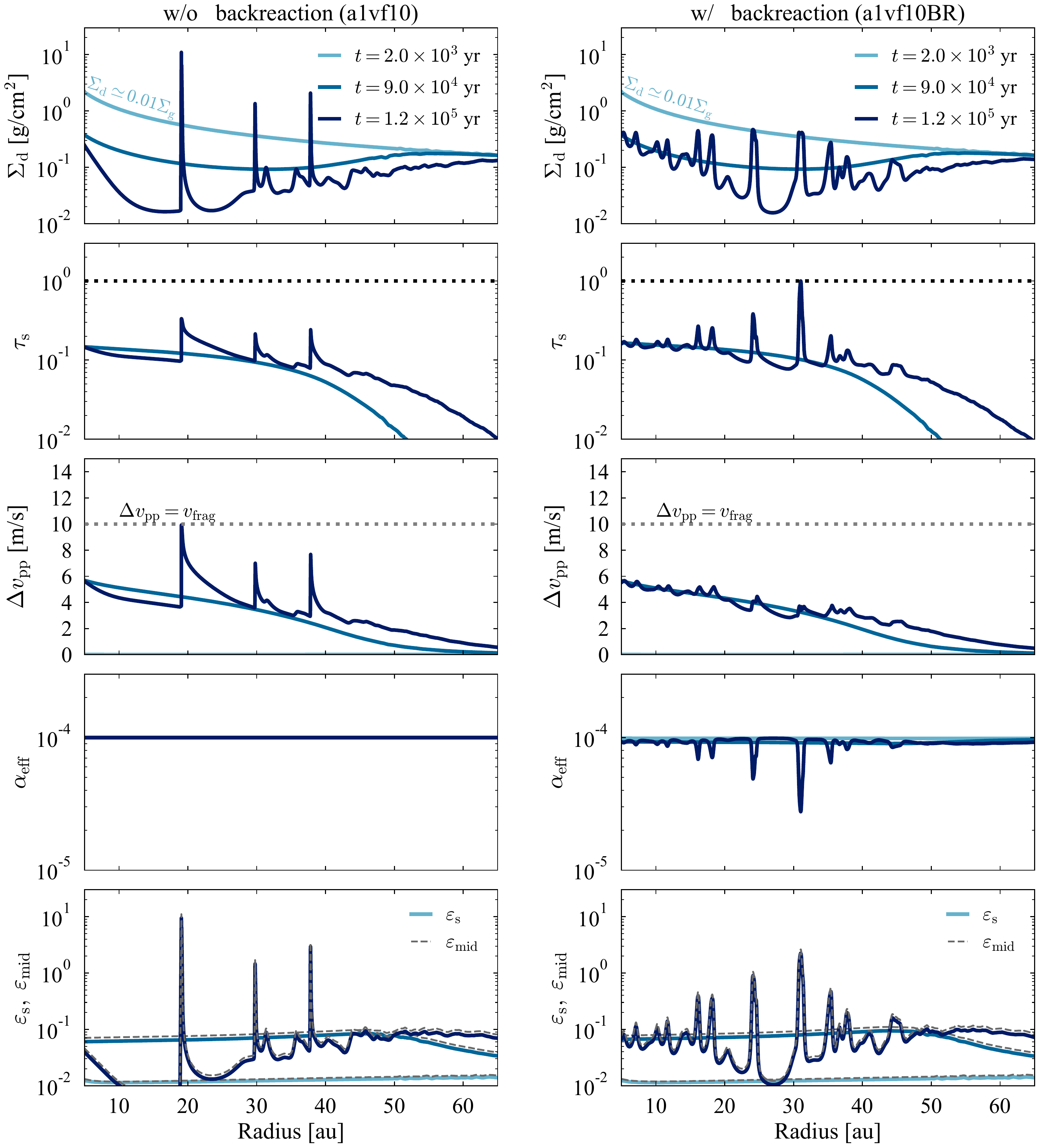} 
	}
	\end{center}
	\vspace{-30pt}
\caption{Radial profiles of dust surface density $\sigmad$ (first row), dimensionless stopping time $\taus$ (second row), the collision velocity $\Delta\vpp$ (third row), the effective turbulence strength $\alpeff$ (fourth row), and the dust-to-gas ratio in the sublayer $\epsub$ (fifth row). We also plot $\epmid$ with a dashed line on the bottom panels. We show the results with and without the effects of the backreaction on the right and left panels, respectively. Different lines on each panel show profiles at different time (see the legend on the first row). Dust grains are not so large yet at $t=2.0\times 10^3$ yr ($\taus\ll 1$, light blue line), and thus the profile of $\sigmad$ at this timestep is a line of $0.01\sigmag$ (see the first row). We note that $\taus$ is less than $10^{-2}$ at $t=2\times 10^3\;\mathrm{yr}$, and thus the light blue line does not appear on the panels in the second row.}
\label{fig:evol_fid}
\end{figure*}

\subsection{Preliminary estimates of collision velocity under the action of backreaction}\label{subsec:preest}
Before analyzing simulation results, we first show how much the backreaction reduces the collision velocity. Figure \ref{fig:ep_vcol_map} shows the collision velocities at $r=30\;\mathrm{au}$ as a function of $\taus$ for $\sigmad/\sigmag=10^{-3},\;10^{-2}$ and $10^{-1.5}$. We adopt a low $\alpha_0$ value ($\alpha_0=1\times 10^{-4}$) since recent ALMA observations suggest weak turbulence \citep[e.g.,][]{Pinte2016,Villenave2022}. In this case, the differential drift velocity dominate the turbulence-induced velocity. The assumed low $\sigmad/\sigmag$ of $10^{-3}$ is motivated by the fact that dust grains get depleted as they grow \citep[e.g.,][]{Brauer2008}. We note that this depletion is efficient when the radial drift is more serious to limit the dust growth than fragmentation. The collision velocity becomes larger than 10 m/s as dust grows in the dust-depleted case (see the black line). However, if dust-to-gas ratio is $1\times10^{-2}$, the maximum collision velocity becomes less than 10 m/s (see the purple line). This means that dust grains can overcome the fragmentation barrier for $\alpha_0=1\times 10^{-4}$ and $\vfrag=10\;\mathrm{m/s}$ if coagulation instability increases $\sigmad/\sigmag$ up to $1\times 10^{-2}$ after the initial dust depletion. On the other hand, dust grains should face the fragmentation barrier for $\sigmad/\sigmag=10^{-2}$ if $\vfrag$ is 3 m/s or smaller. More significant dust enrichment is necessary in such a case (see the orange line). Dust grains at inner hot region should need further enhancement since the turbulent collision velocity increases as the temperature increases \citep[e.g.,][]{Weidenschilling1993,Ormel2007,Birnstiel2012,Drazkowska2016}. In this way, the maximum dust-to-gas ratio achieved via coagulation instability is the key physical value to figure out whether or not dust grains avoid fragmentation. 

The collision velocity becomes larger for larger $\alpha_0$ \citep[e.g.,][]{Weidenschilling1993,Ormel2007}. Thus, more significant dust enrichment via coagulation instability is necessary for dust to overcome the fragmentation in more turbulent disks. We explore the $\alpha_0$-dependence of the nonlinear development of coagulation instability as well as $\vfrag$-dependence in Section \ref{subsec:params} (Figure \ref{fig:summary_param}; see also Section \ref{subsec:vfrag}).

\begin{figure}[tp]
	\begin{center}
	\hspace{100pt}\raisebox{20pt}{
	\includegraphics[width=1.0\columnwidth]{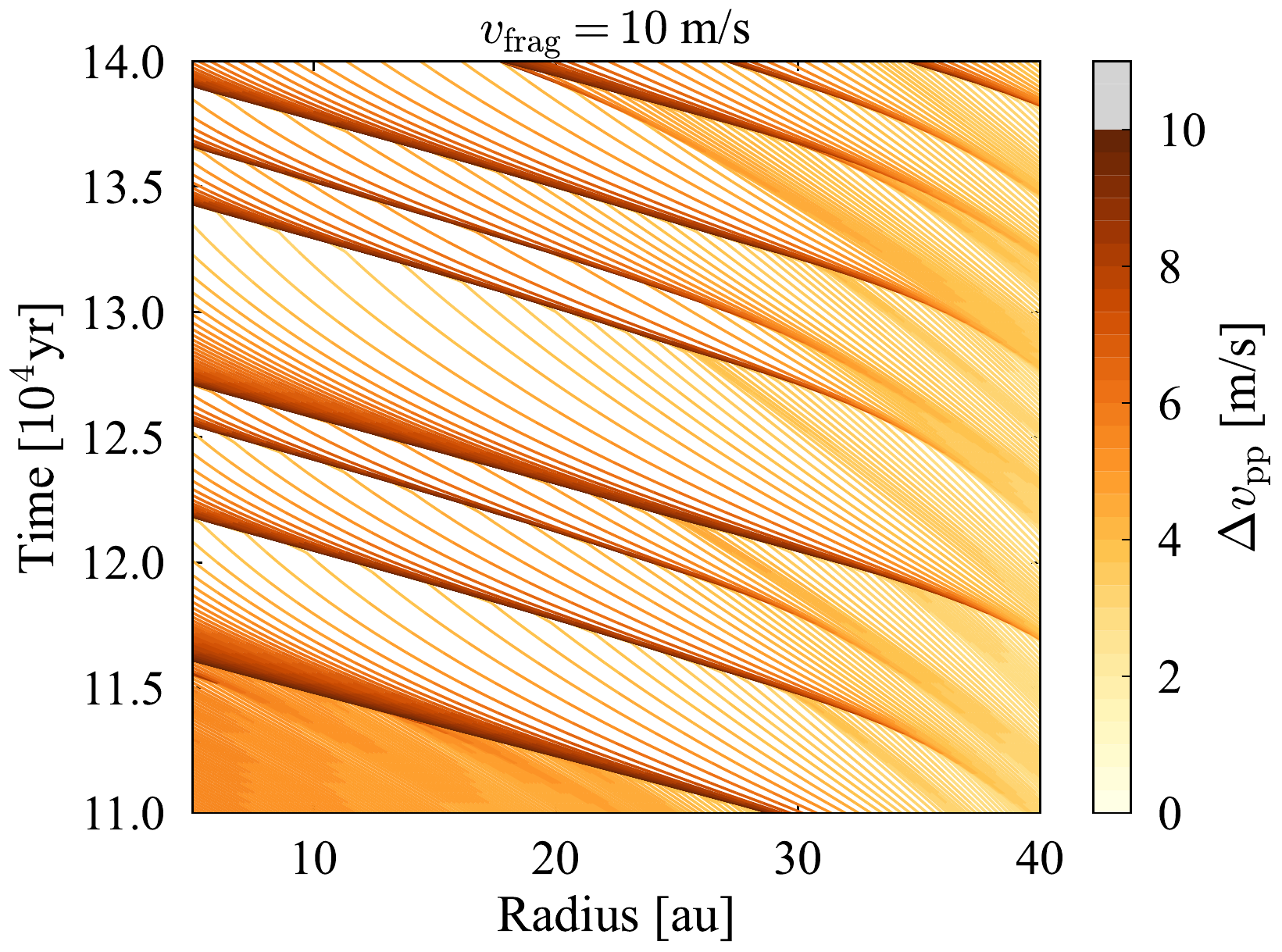}
	}
	\end{center}
	\vspace{-30pt}
\caption{The dust collision velocity $\Delta\vpp$ on the $r-t$ plane obtained from the a1vf10 run. Each colored line show trajectory of each dust cell boundary. Color of the lines represents the magnitude of $\Delta\vpp$. We arrange the line color so that a line is highlighted in lightgrey if $\Delta\vpp$ is larger than 10 m/s. Since our method is based on the Lagrangian cells, the dust surface density is larger at radii where more lines concentrate. We note that we used a reduced number of dust cells to plot this figure. }
\label{fig:dvtot_on_r-t}
\end{figure}

\begin{figure}[tp]
	\begin{center}
	\hspace{100pt}\raisebox{20pt}{
	\includegraphics[width=0.9\columnwidth]{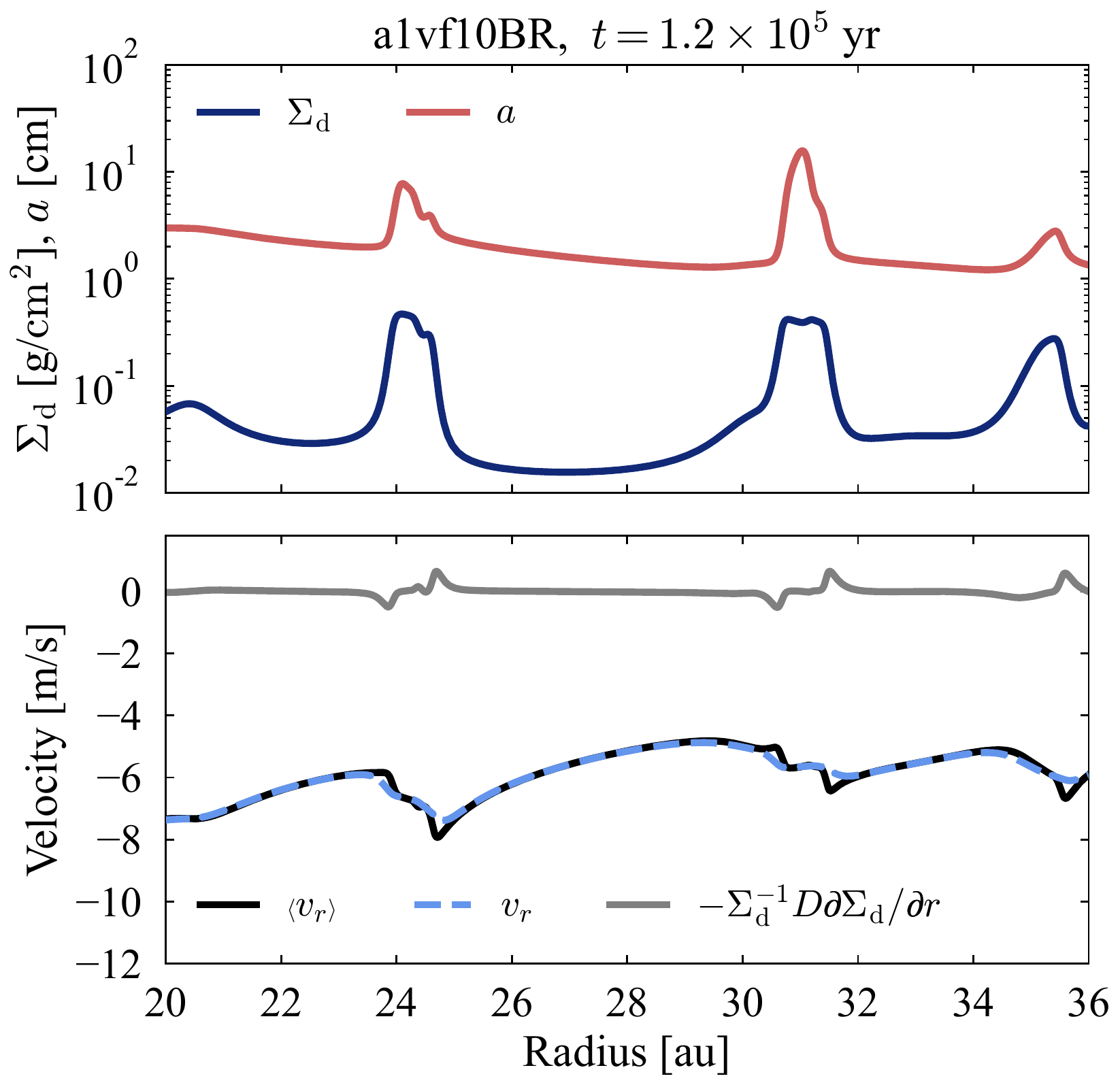}
	}
	\end{center}
	\vspace{-30pt}
\caption{The resulting structures in a1vf10BR run. (Top panel) The radial profiles of dust surface density (dark blue line) and dust size $a$ (red line) are shown. We find that the sufficiently evolved ring shows a plateau structure in the dust surface density profile. This is in contrast to the result of a1vf10 run (the left panels of Figure \ref{fig:evol_fid}) and of Paper I. The dust sizes increases locally up to the size of $\taus=1$, which is consistent the result in Paper I. (Bottom panel) The radial profiles of the mean drift velocity $\left<v_r\right>$ (black line), the total velocity $v_r$ (blue dashed line), and the diffusion velocity (gray line) are shown. }
\label{fig:ring_prop}
\end{figure}

\section{Results of simulations}\label{sec:results}

\subsection{Fiducial case}\label{subsec:fiducial}

We show the results of the a1vf10BR run as a fiducial case. The right five panels in Figure \ref{fig:evol_fid} show the time evolution of the radial profiles of the dust surface density $\sigmad$, the dimensionless stopping time $\taus$, the collision velocity $\Delta\vpp$, the effective turbulence strength $\alpeff$, and the dust-to-gas ratio in the sublayer $\epsub$. For comparison, we also conducted a simulation without the backreaction (a1vf10 run), where we ignore the $\varepsilon$-dependence of both $\alpha$ and the drift velocities (Equations (\ref{eq:alpeff}), (\ref{eq:vrdrift}), (\ref{eq:vrcol}) and (\ref{eq:vphicol})). The results of a1vf10 run are shown on the left panels in Figure \ref{fig:evol_fid}.

We first overview the difference between the a1vf10 run and the a1vf10BR run seen in Figure \ref{fig:evol_fid} before explaining the results in detail. The first row shows the surface density profiles. The rings seen in the a1vf10 run are narrower than the rings in the a1vf10BR run. The second row shows that $\taus$ in the ring is smaller than unity in the a1vf10 run while $\taus$ reaches unity in the a1vf10BR run. This difference comes from the difference in the collision velocity. The large collision velocity leads to efficient fragmentation and the limited dust growth in the a1vf10 run, i.e., $\Delta \vpp\simeq\vfrag$ (the third row, see also Figure \ref{fig:dvtot_on_r-t}). On the other hand, the collision velocity is kept smaller than $\vfrag$ because of the backreaction in the a1vf10BR run. The backreaction also reduces the turbulence strength locally in the a1vf10BR run (the fourth row). Since we switch off the backreaction in the a1vf10 run (the left column), the turbulence strength does not change regardless of high dust-to-gas ratio (e.g., for $\epsub$ see the bottom row). 

Next, we describe the resulting structures and those differences due to the backreaction in more detail. In the absence of the backreaction (the left panels of Figure \ref{fig:evol_fid}), the nonlinear development of coagulation instability creates dense and narrow dust rings at $r\simeq20,\;30,\;40\;\mathrm{au}$. The most developed ring has $\sigmad/\sigmag\sim0.1$ and $\epsub\sim10$ (see the first and fifth rows of Figure \ref{fig:evol_fid}). We note that the dust grains are not decelerated regardless of large $\epsub$ since we switch off the backreaction in the a1vf10 run. The ring formation caused by coagulation instability was also found in Paper I. We newly find that collisional fragmentation affects nonlinear growth of coagulation instability. Fragmentation reduces the growth efficiency of coagulation instability for $\Delta\vpp>\vfrag/5=2\;\mathrm{m/s}$ (see Equation (\ref{eq:peff})). The nonlinear growth is saturated once the collision velocity is equal to $\simeq\vfrag$ (see the peak at $r\simeq 20\;\mathrm{au}$ of the third panel on the left column of Figure \ref{fig:evol_fid}). Two rings at $r\simeq30\;\mathrm{au}$ and $40\;\mathrm{au}$ are in the middle of development and do not reach the saturation state at $t\simeq1.2\times 10^5$ yr. The later evolution of those rings can be seen in Figure \ref{fig:dvtot_on_r-t}, where we plot $\Delta\vpp$ at each dust cell on the space-time plane. The colored lines show the trajectories of dust cell boundaries. Since our method utilizes the Lagrangian cells, the trajectory of one ring is seen as a bundle of the lines. As in the linear growth phase \citep[see][]{Tominaga2021,Tominaga2022}, we can see that the rings move faster than dust grains and sweep inner dust grains.  Along the bundles, the collision velocities first increase but never become larger than $\vfrag=10\;\mathrm{m/s}$. Thus, dust grains in the well-developed dust rings have so-called fragmentation-limited sizes when the nonlinear coagulation instability is saturated. The fragmentation-limited $\taus$ is smaller than unity but relatively large ($>0.1$) in the present case. As a result, the resulting rings keep drifting toward the central star.

The resulting structures are different in the presence of the backreaction (the right panels of Figure \ref{fig:evol_fid}). First, the resulting rings are wider than the rings without the backreaction (see also Paper I), and the dust surface density profile shows plateau structures in the rings. The top panel of Figure \ref{fig:ring_prop} shows the dust surface density profile $\sigmad$ (the dark blue line) and the dust size profile $a$ (the red line) around the rings. The rings at $r\simeq24\;\mathrm{au}$ and $31\;\mathrm{au}$ show plateau profiles, which are in contrast to the simulation result without the backreaction. Interestingly, the center of the plateau structure in $\sigmad$ roughly coincides with the peak of the dust size profile. This is in contrast to the results without the backreaction in Paper I, in which we find that the peak of $\sigmad$ is offset from the peak of $a$ even in the nonlinear growth phase. We attribute these profiles to transition from coagulation instability to pure coagulation. After the dust concentration leads to $\epsub\sim1$ (see the fifth row of Figure \ref{fig:evol_fid}), the drift deceleration due to the backreaction reduces the growth efficiency of coagulation instability \citep[][]{Tominaga2021}. In other words, the density increase due to the instability becomes insignificant. On the other hand, the dust size keeps increasing via pure coagulation toward the size of $\taus=1$. The larger dust-to-gas surface density ratio ($\sim0.01$, see the top panel of Figure \ref{fig:evol_fid}) resulting from the nonlinear coagulation instability enables the faster size growth than in the background with the low dust-to-gas ratio ($\sim10^{-3}$).

Second, the drift velocity of the ring is reduced by the backreaction (Equation (\ref{eq:vrdrift})). This saves the dust ring from flowing out of the numerical domain, which is in contrast to the a1vf10 run. The bottom panel of Figure \ref{fig:ring_prop} shows the radial profiles of $\left<v_r\right>$ (the black line), $v_r$ (the blue dashed line) and the diffusion velocity $-\sigmad^{-1}D\partial\sigmad/\partial r$ (the gray line). The diffusion velocity is small in the rings because of the plateau structure in the $\sigmad$ profile. The drift speed of the ring at $r\simeq31\;\mathrm{au}$ is $\simeq 6\;\mathrm{m/s}\simeq 1\times10^{-3}\;\mathrm{au/yr}$. This speed is only 20 percents of the maximum drift speed $\eta\vk$, which is explained by the large dust-to-gas ratio in the ring at the final time step ($\epsub\simeq 2$). Further dust growth and sedimentation will reduce the drift velocity more.

\begin{figure}[tp]
	\begin{center}
	\hspace{100pt}\raisebox{20pt}{
	\includegraphics[width=0.9\columnwidth]{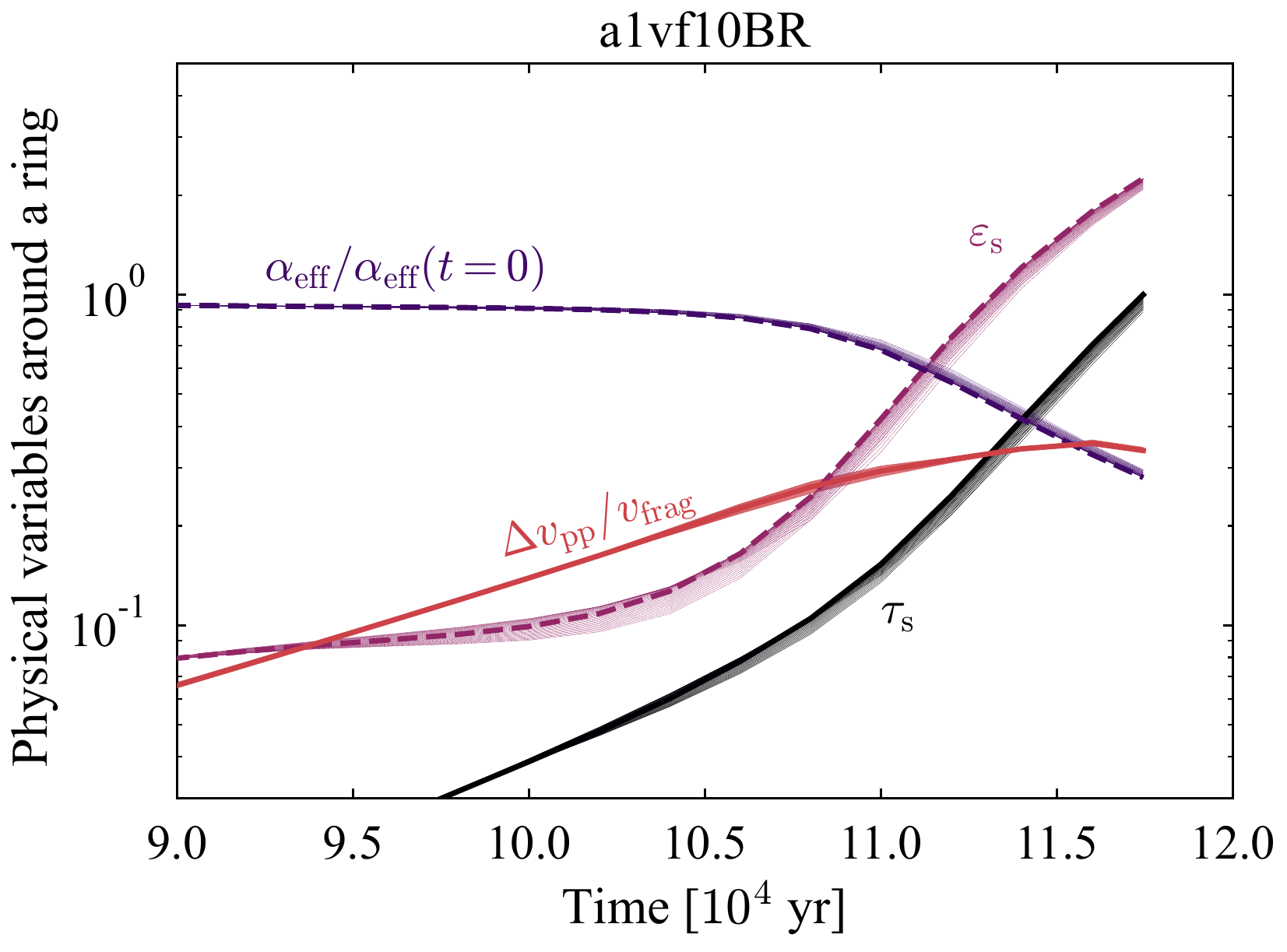} 
	}
	\end{center}
	\vspace{-30pt}
\caption{Time evolution of $\taus$ (black solid line), $\Delta \vpp/\vfrag$ (red solid line), $\epsub$ (purple dashed line), and $\alpeff/\alpeff(t=0)$ (dark purple dashed line) of the dust cell whose $\taus$ reaches unity at the final time step. The thin lines correspond to the evolution of dust cells of $i_{\mathrm{ring}}-15\leq i\leq i_{\mathrm{ring}}+15$, where $i$ is the cell number and $i_{\mathrm{ring}}$ is the cell number of the dust cell whose $\taus$ reaches unity at the final time step.}
\label{fig:tevol_various}
\end{figure}

\begin{figure}[tp]
	\begin{center}
	\hspace{100pt}\raisebox{20pt}{
	\includegraphics[width=0.9\columnwidth]{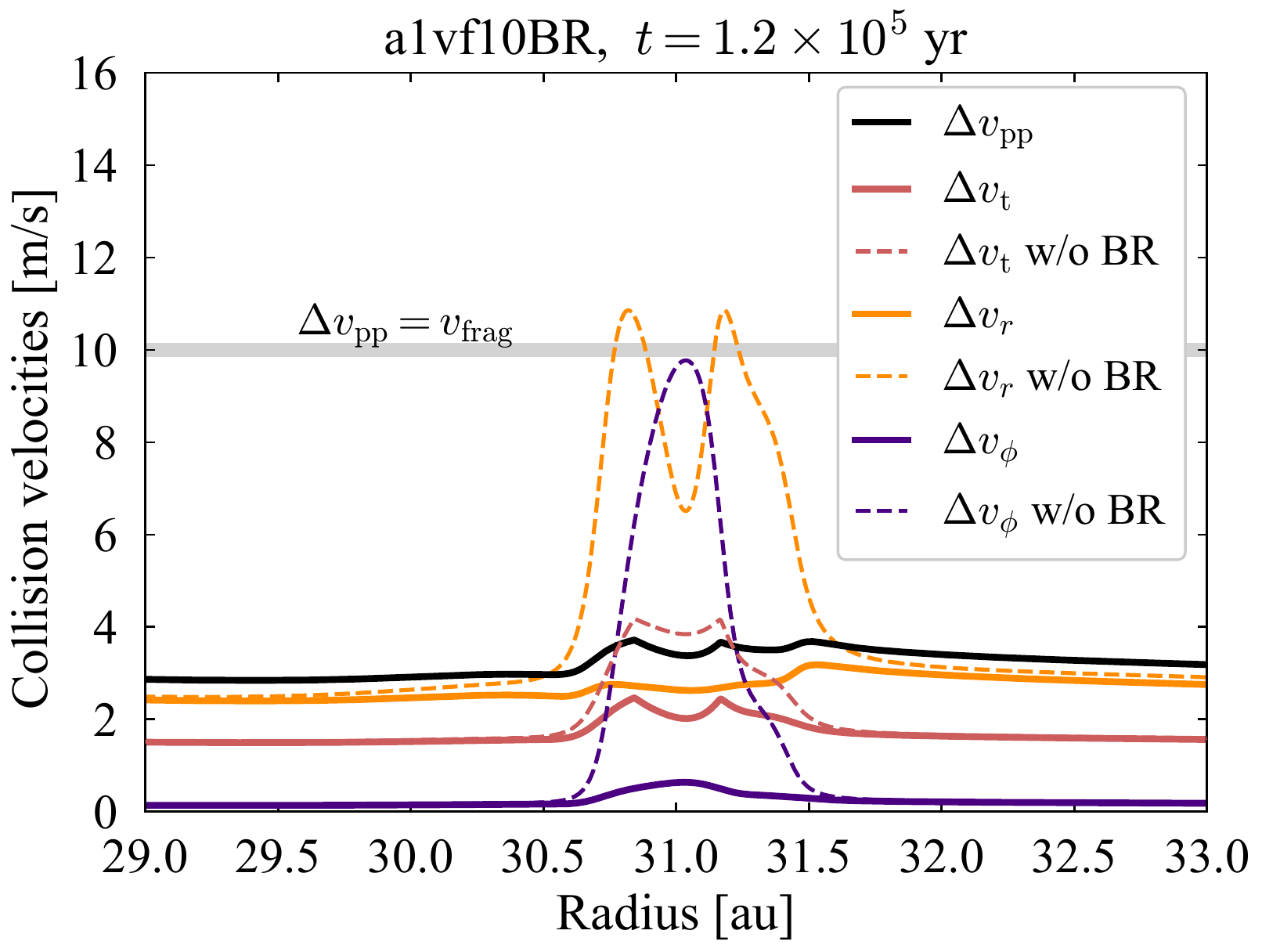}
	}
	\end{center}
	\vspace{-30pt}
\caption{The radial profiles of the collision velocities due to the turbulent motion (red lines), the radial drift (orange lines), and the azimuthal drift (purple lines) at the final time step. The black line is the total collision velocity. The solid lines show the collision velocities in the presence of the backreaction while the dashed lines show those in the absence of the backreaction. The dust grains are faced with the fragmentation if the colored or black lines cross the gray line. We can see that the backreaction reduces the collision velocities in the ring ($r\simeq 31\;\mathrm{au}$), and the total collision velocity becomes less than $\vfrag$.}
\label{fig:collvel}
\end{figure}

Third, we find the collision velocity is smaller than $\vfrag$ even at the nonlinear growth phase (see the third panel on the right column in Figure \ref{fig:evol_fid}). This is also due to the backreaction. Figure \ref{fig:tevol_various} shows the time evolution of $\taus$ (black solid line), $\Delta \vpp/\vfrag$ (red solid line), $\epsub$ (purple dashed line), and $\alpeff/\alpeff(t=0)$ (dark purple dashed line) of the dust cell whose $\taus$ reaches unity at the final time step. We use $i_{\mathrm{ring}}$ below to denote a cell number of this dust cell. We also plot each physical value of the $i$th cell around the $i_{\mathrm{ring}}$th cell for $i_{\mathrm{ring}}-15\leq i\leq i_{\mathrm{ring}}+15$ with thin lines. The ratio $\Delta \vpp/\vfrag$ increases as $\taus$ increases, but the increasing rate becomes smaller for $t\gtrsim1.1\times10^5\;\mathrm{yr}$ even though $\taus$ keeps increasing. The resulting reduced $\Delta \vpp$ is due to both large dust-to-gas ratio in the sublayer, i.e. $\epsub\gtrsim1$ (see the fifth row of Figure \ref{fig:evol_fid}), and the decrease in $\alpeff$ due to the backreaction with $\epmid\gtrsim1$ (see also Figure \ref{fig:epmid_dependence}).

Figure \ref{fig:collvel} shows the radial profiles of the collision velocities around the most collapsed ring ($r\simeq31\;\mathrm{au}$). The figure shows that the total collision velocity in the presence of the backreaction is smaller than $\vfrag$, meaning that dust grains avoid catastrophic fragmentation in the ring. The colored solid (dashed) lines in Figure \ref{fig:collvel} show each component of the collision velocity in the presence (absence) of the backreaction. The differential drift velocities are reduced most significantly in the a1vf10BR run, which helps dust grains avoid the fragmentation. The reduction of the turbulence-induced collision velocity is subdominant in this run compared to the reduction of the differential drift velocities.

\begin{figure}[tp]
	\begin{center}
	\hspace{100pt}\raisebox{20pt}{
	\includegraphics[width=0.9\columnwidth]{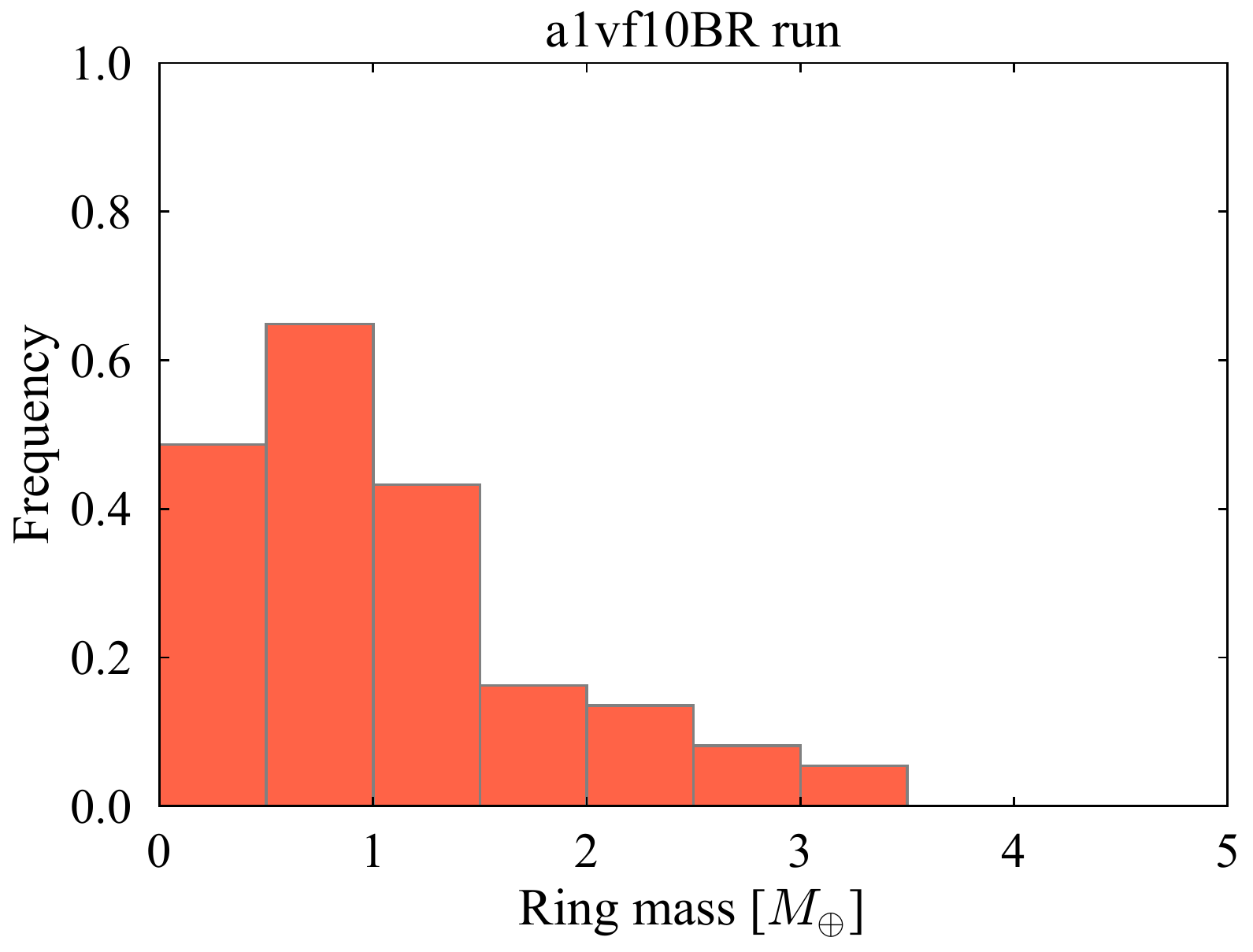}
	}
	\end{center}
	\vspace{-30pt}
\caption{Histogram of the ring mass from a1vf10BR run. We conduct another five runs for the same parameter but with different phase of initial perturbations and compile the results. The ring mass is defined as dust masses within the full width at half maximum of the dust surface density.}
\label{fig:ringmass_a1vf10BR}
\end{figure}

We finally evaluate a ring mass. We derive the ring mass by summing up dust-cell masses within the full width at half maximum (FWHM) of $\sigmad$. We merge two rings and count them as a single ring if the difference in $\sigmad$ between the local maximum and the local minimum is within 20 percent \footnote{We select a ring whose ring-gap contrast is larger than 50 percent to exclude weak linear perturbations in the dust surface density profile as much as possible.}. To eliminate the dependence on the initial perturbations, we conduct another five runs with different phase of perturbations and derive a ring-mass histogram. Figure \ref{fig:ringmass_a1vf10BR} shows the result. We find that the ring mass of $\simeq 0.5\merth-1\merth$ is the most frequent. The mass of a single ring is much smaller than the initial total dust mass ($\simeq2.4\times10^2\merth$). Therefore, forming multiple rings will result in a large amount of planetesimals.

\begin{figure*}[htp]
	\begin{center}
	\hspace{100pt}\raisebox{20pt}{
	\includegraphics[width=1.9\columnwidth]{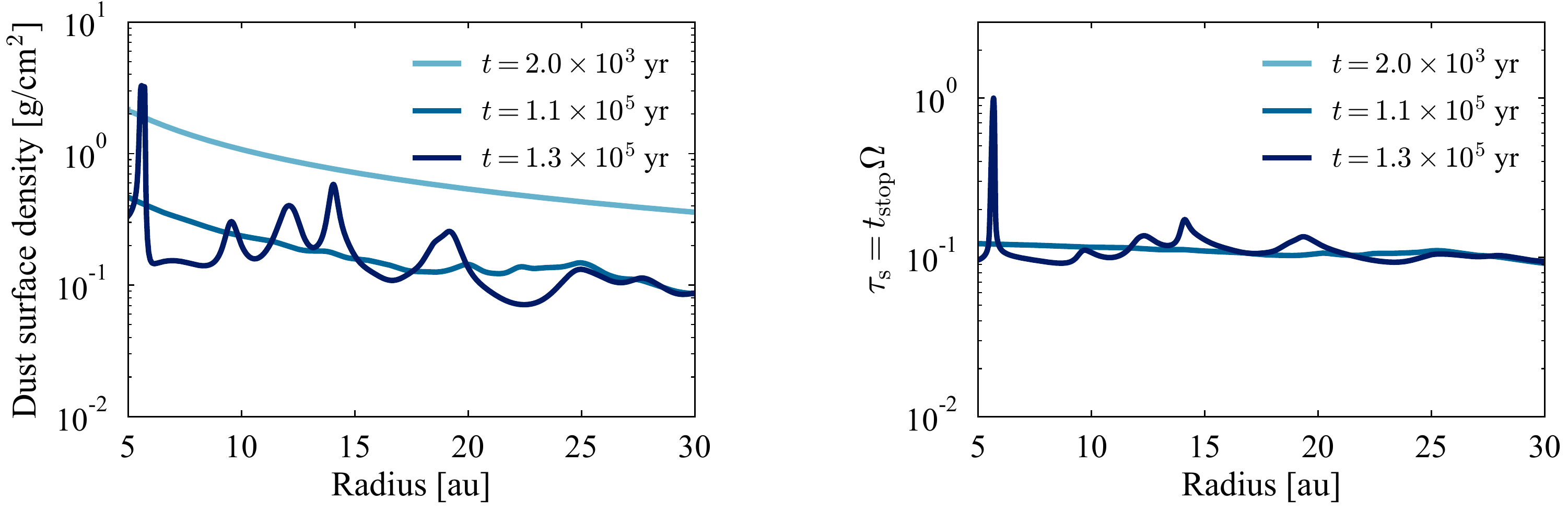}
	}
	\end{center}
	\vspace{-30pt}
\caption{The results of a3vf10BR run. The left and right panels show the time evolution of $\sigmad$ and $\taus$. The densest dust ring forms at inner radii compared to a1vf10BR run (see also Appendix \ref{app:on_a3vf10br}). This is due to large diffusion that reduces the growth rate of coagulation instability. The dust size increases up to the size of $\taus=1$ in the resulting ring. We note that $\taus$ is less than $10^{-2}$ at $t=2\times 10^3\;\mathrm{yr}$, and thus the light blue line does not appear on the right panel.}
\label{fig:a3vf10BR}
\end{figure*}

\begin{figure}[htp]
	\begin{center}
	\hspace{100pt}\raisebox{20pt}{
	\includegraphics[width=0.9\columnwidth]{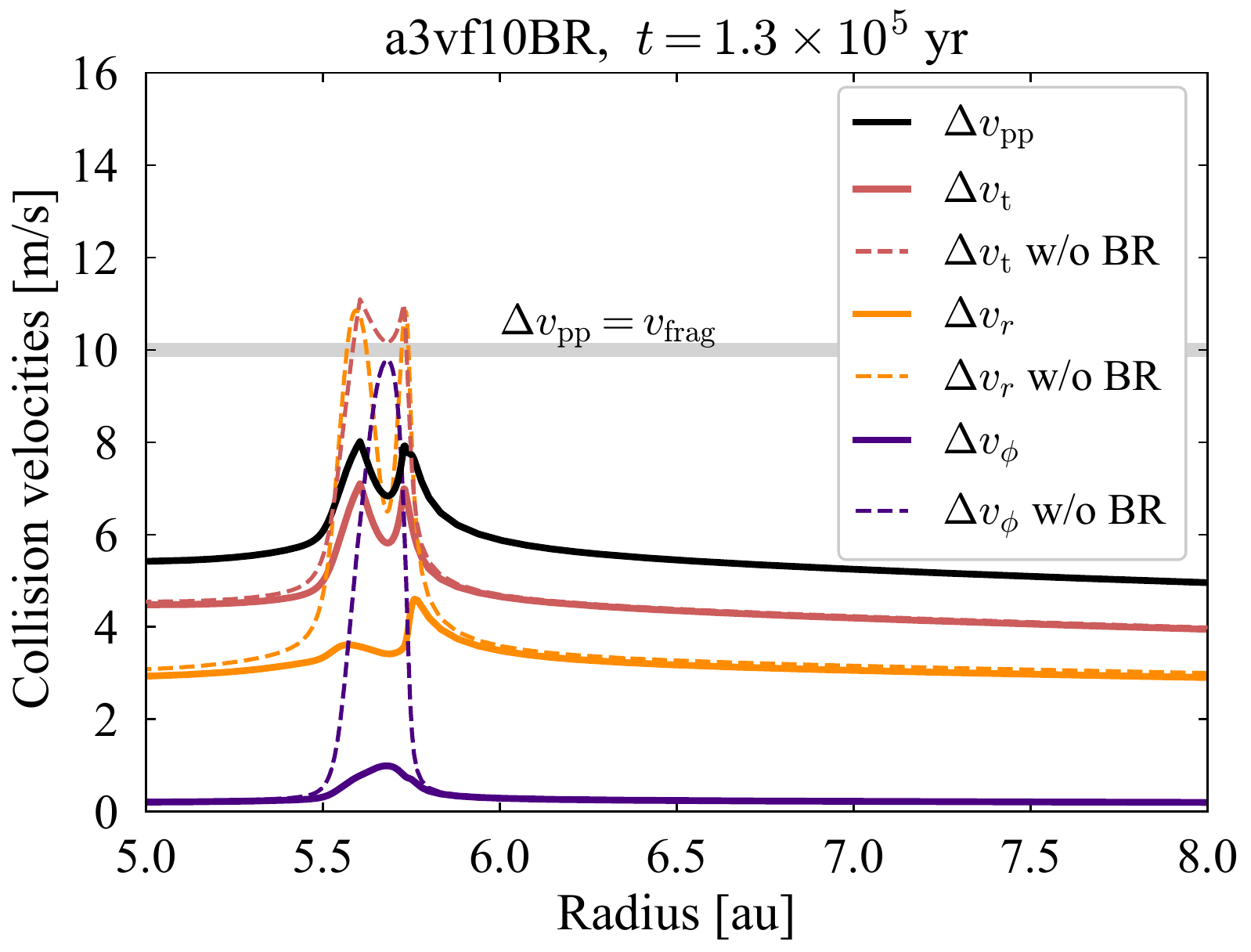}
	}
	\end{center}
	\vspace{-30pt}
\caption{The radial profiles of the collision velocities at the final time step obtained in a3vf10BR run . The colored solid (dashed) lines represent each collision velocity component in the presence (absence) of the backreaction as in Figure \ref{fig:collvel}. The dominant velocity component is the turbulence-induced velocity in a3vf10BR run (see the red lines). We can see that the backreaction reduces the collision velocities in the ring, and the total collision velocity becomes less than $\vfrag$.}
\label{fig:collvel_a3vf10BR}
\end{figure}

\begin{figure*}[htp]
	\begin{center}
	\hspace{100pt}\raisebox{20pt}{
	\includegraphics[width=1.8\columnwidth]{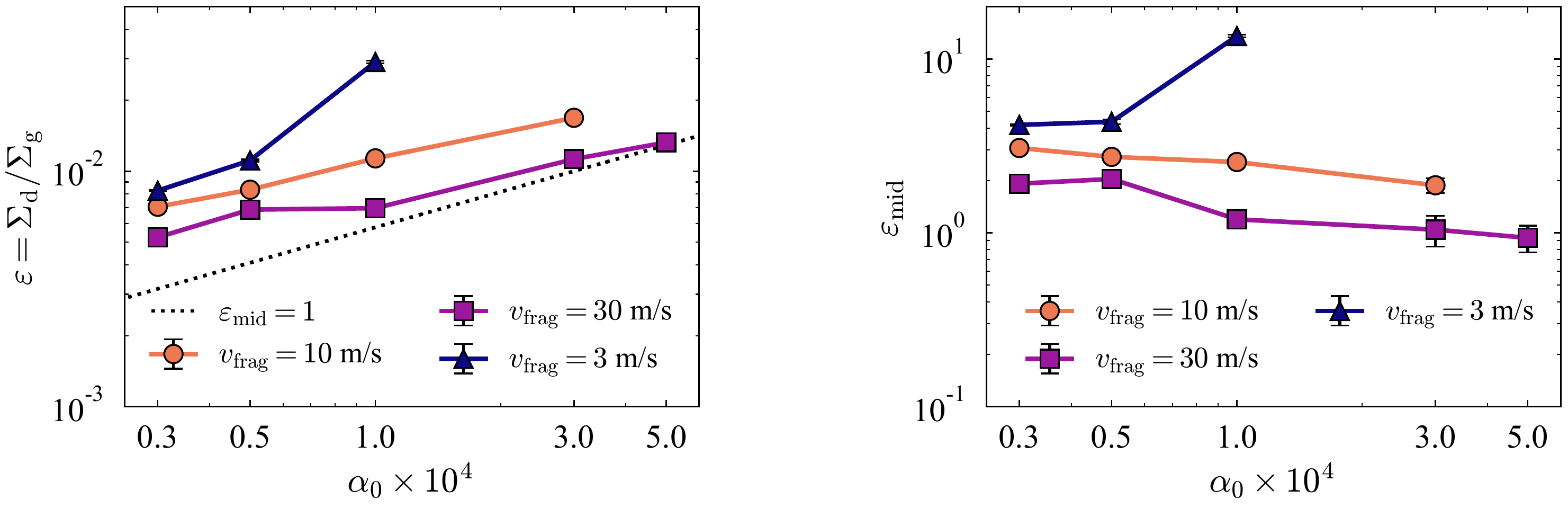}
	}
	\end{center}
	\vspace{-30pt}
\caption{The dust-to-gas surface density ratio $\varepsilon$ (left panel) and the dust-to-gas ratio at the midplane (right panel) around the dust cell ($i_{\mathrm{ring}}$th cell) whose $\taus$ reaches unity at the final timestep. The marks show the averaged values for $i_{\mathrm{ring}}-15\leq i \leq i_{\mathrm{ring}}+15$ where $i$ is the cell number. The error bars correspond to the standard deviation. The surface density ratio slightly increases as $\alpha_0$ while the $\epmid$ is roughly constant. The black dotted line on the left panel shows the dust-to-gas surface density ratio that we semi-analytically derive using $\epmid=1$ and $\taus=1$. The black dotted line roughly reproduces the $\alpha_0$-dependence of $\varepsilon$.} 
\label{fig:maximumDG}
\end{figure*}

In summary, we observe the following in the a1vf10 run and the a1vf10BR run: 
\begin{itemize}
\item Nonlinear coagulation instability creates narrow rings and is saturated by the fragmentation in the absence of the backrection.
\item Dust rings are wider in the presence of the backreaction, and the dust surface density shows a plateau structure in the resulting ring.
\item The backreaction reduces the collision velocities and allows dust growth up to $\taus=1$ in the resulting ring.
\item Since the drift velocity is reduced, dust grains will be retained through further evolution beyond $\taus=1$.
\item Rings with a mass of $0.5\merth-1\merth$ are the most frequent.
\end{itemize}

\subsection{Parameter dependence}\label{subsec:params}

\begin{figure}[htp]
	\begin{center}
	\hspace{0pt}\raisebox{20pt}{
	\includegraphics[width=0.9\columnwidth]{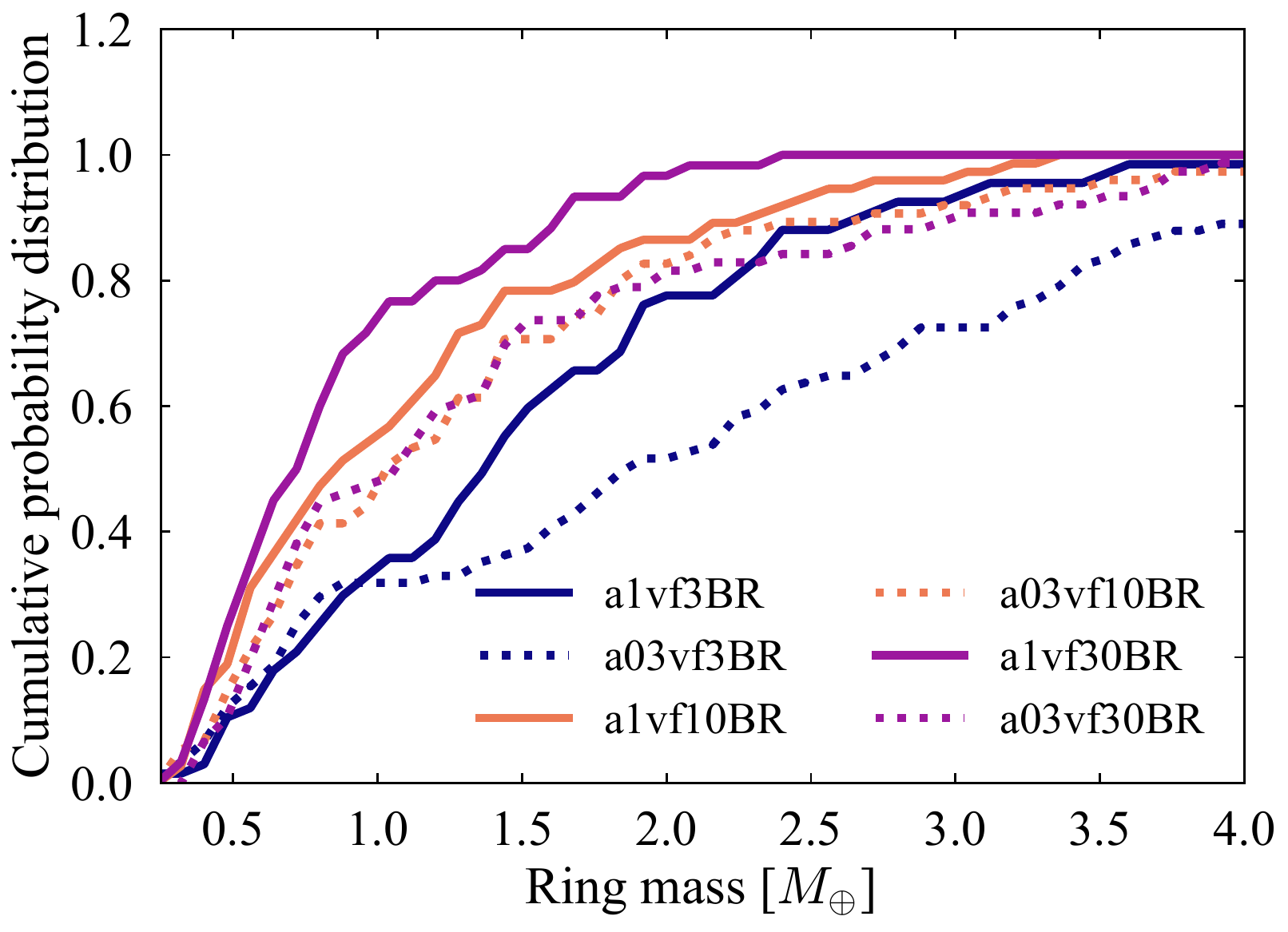}
	}
	\end{center}
	\vspace{-30pt}
\caption{Cumulative probability distribution of the ring mass in 6 runs. We conduct another five runs for each parameter set with different initial perturbations to plot these distributions. In most runs, we find that a ring mass of $\simeq0.5\merth-1.5\merth$ is the most frequent.}
\label{fig:mass_cdf_param}
\end{figure}

Next we show results of the parameter studies. Figure \ref{fig:a3vf10BR} shows the time evolution of $\sigmad$- and $\taus$-profiles obtained in the a3vf10BR run, where we use a value of $\alpha_0$ three times larger than that in the a1vf10BR run. Coagulation instability grows more slowly for stronger radial diffusion and creates dense rings at inner radii as shown in Paper I.\footnote{The densest ring consists of dust cells that are initially located around the inner boundary of the perturbed region, which is also similar to the results in Paper I.} Regardless of the slow growth, the nonlinear coagulation instability accelerates dust growth and leads to $\taus=1$ as in the a1vf10BR run before the ring flows out of the numerical domain. Although the most developed ring forms near the inner boundary, we confirm that this is physical and not due to a boundary effect (see Appendix \ref{app:on_a3vf10br}). We plot the radial profiles of the collision velocities in Figure \ref{fig:collvel_a3vf10BR} as in Figure \ref{fig:collvel}. The turbulence-induced collision velocity is the largest velocity. As observed in the a1vf10BR run, the backreaction decreases the collision velocities, and the total collision velocity (the black line) becomes smaller than $\vfrag$. This reduction of the collision velocities allows dust growth toward $\taus=1$ in the ring without catastrophic fragmentation as in the a1vf10BR run.

We find that the ring in the nonlinear regime is slightly narrower for larger $\alpha_0$. As a result, the dust-to-gas surface density ratio $\varepsilon$ in the ring at the final time step ($\taus=1$) increases with $\alpha_0$ (see the left panel of Figure \ref{fig:maximumDG}). On the other hand, the dust-to-gas ratios at the midplane and in the sublayer are roughly independent of $\alpha_0$ for a given $\vfrag$ (see the right panel of Figure \ref{fig:maximumDG} for $\epmid$).\footnote{The difference between $\epmid$ and $\epsub$ is less than $\simeq2$ at the final timestep.} When $\alpha_0$ is large, the vertical stirring is efficient and the dust-to-gas ratio in the dust layer becomes smaller for a given surface density ratio $\varepsilon$. In other words, the dust drift is less reduced for larger $\alpha_0$. As found in \cite{Tominaga2021}, coagulation instability grows more efficiently for weaker backreaction (see Section 4.2 therein). Thus, larger $\alpha_0$ leads to the greater enhancement of the surface density ratio $\varepsilon$ although the stronger ``radial" diffusion delays the linear growth of the instability. The radial concentration via coagulation instability proceeds until the dust-to-gas ratios at the midplane and in the sublayer become comparable to unity. We estimate the saturated value of $\varepsilon$ under the assumption of $\epmid=1$ and $\taus=1$ as follows:
\begin{equation}
\varepsilon(\epmid=1)\simeq 5.8\times10^{-3}\left(\frac{\alpha_0}{1\times10^{-4}}\right)^{1/2}.\label{eq:ep_sat}
\end{equation}
Equation (\ref{eq:ep_sat}) roughly reproduces the $\alpha_0$-dependence of $\varepsilon$ seen in the simulation results (see the black dotted line on the left panel of Figure \ref{fig:maximumDG}).

We also find that $\varepsilon$, $\epmid$, and $\epsub$ at the final time step ($\taus=1$) are slightly larger for smaller $\vfrag$. This is because a large dust-to-gas ratio is required for small $\vfrag$ to avoid fragmentation (see Figure \ref{fig:ep_vcol_map}).

\begin{figure}[tp]
	\begin{center}
	\hspace{0pt}\raisebox{20pt}{
	\includegraphics[width=0.9\columnwidth]{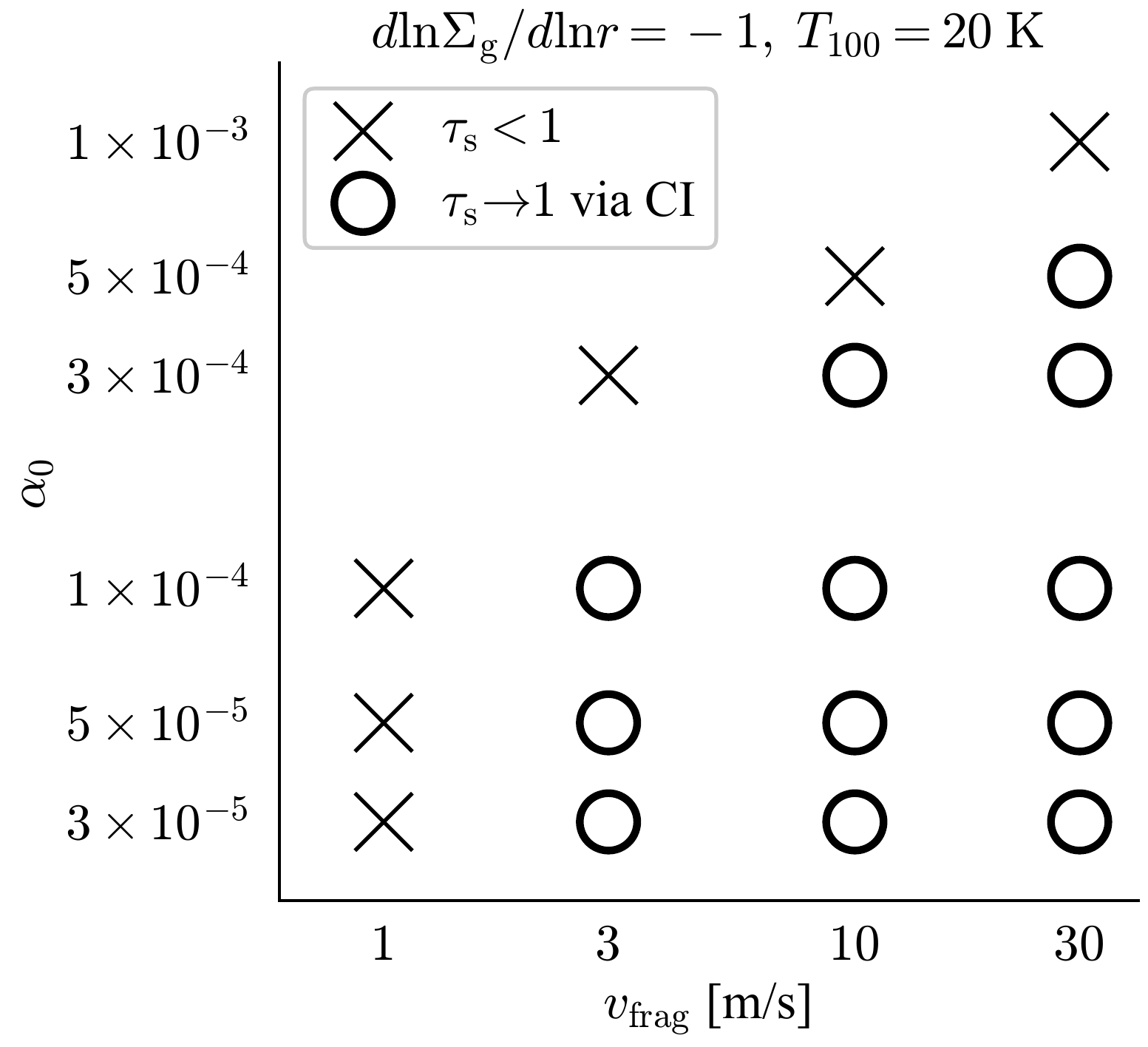}
	}
	\end{center}
	\vspace{-30pt}
\caption{Summary of the parameter study. The open circles mean that coagulation instability (CI) develops, radially concentrates dust grains, and leads to $\taus\to1$ in the resulting ring. The cross marks mean that the instability is inefficient and does not increase $\taus$ up to unity because of too small $\vfrag$ or because of too large $\alpha_0$ that reduces the growth rate of the instability as found in Paper I. 
}
\label{fig:summary_param}
\end{figure}

Figure \ref{fig:mass_cdf_param} shows the cumulative probability distribution of a ring mass obtained from 6 runs. The most frequent ring mass is about $\simeq0.5\merth-1.5\merth$ in most runs, which is similar to the result of the a1vf10BR run. Since the total dust mass of the initially perturbed region is $\simeq1.2\times10^2\merth$, we may expect that at most $\sim50-160$ rings\footnote{The obtained ring mass is dust mass within FWHM of $\sigmad$. Thus, the mass between the centers of two gaps adjacent to one ring should be $\simeq0.75\merth-2.25\merth$. Taking this factor into account, we evaluate the total number of rings.} form in total if all dust grain initially located in the perturbed region is trapped in rings. The frequency of massive rings (a few $\merth$) is higher in the a03vf3BR run than in the other runs. We find that these relatively massive rings form via the merging of rings. The ring merging reduces the total number of formed rings. How many rings remain in the disk in the end is an issue to be addressed.

Figure \ref{fig:summary_param} summarizes the results of the parameter study. The open circles mean that the nonlinear coagulation instability operates, radially concentrates dust grains, and leads to $\taus\to1$ in the resulting ring. The cross marks mean that the instability is inefficient and does not increase $\taus$ up to unity. As mentioned above and in Paper I, strong dust diffusion delays the growth of coagulation instability and prevents formation of dense rings and the dust-growth acceleration. The fragmentation also delays the instability \citep[see Section 4.5 in][]{Tominaga2021}. Coagulation instability does not grow when the fragmentation velocity is too small and the background dust size is limited by fragmentation. In this way, small $\alpha_0$ and large $\vfrag$ are preferable for coagulation instability to efficiently develop. Figure \ref{fig:summary_param} shows that the necessary condition is $\alpha_0< 1\times 10^{-3}$ and $\vfrag> 1\;\mathrm{m/s}$.

\section{Discussions}\label{sec:discussion}

\subsection{Comparison with Paper I}\label{subsec:comp_paperI}
In Paper I where we neglect the backreaction, we find that the ring drift speed $|v_r|$ is larger than the drift speed of dust in the adjacent inner gap. The present simulations with the backreaction also show that the dust in the ring move slightly faster than dust in the gap (see Figure \ref{fig:ring_prop}). The reason why the drift of the ring is faster is the following. The dust-to-gas ratio in the sublayer $\epsub$ is not yet so large and $\simeq2$ at the final time step ($t=1.2\times10^5\;\mathrm{yr}$). On the other hand, the difference in $\taus$ between the dust in the ring and the dust in the gap is $\simeq10$. As a result, the drift speed of the dust grains in the ring is higher than that of the dust grains in the gap although the backreaction decreases their velocity difference (cf. Figure 6 in Paper I).

We estimated the collision probability between a dust grain in the ring and a drifting dust grain in Paper I and showed that the collision was inefficient especially after the dust in the ring grows much larger and $\taus$ becomes larger than unity (see Equation (30) in Paper I). This holds for most of the present results, e.g. the a1vf10BR run. According to Equation (30) in Paper I, the collision probability is less than unity for the ring at $r\simeq31\;\mathrm{au}$ whose radial width is $\simeq0.5-1\;\mathrm{au}$ and $\varepsilon$ is $\simeq0.01$. The dust ring and the drifting dust become more collisionless as the dust growth proceeds in the ring and $\taus$ in the ring increases. Thus, the drifting dust grains possibly pass through the rings even in the present case.

\subsection{On the critical fragmentation velocity and turbulence strength}\label{subsec:vfrag}
According to the equal-mass collision simulations in \citet{Wada2009}, the fragmentation velocities of water ice with $0.1\;\mu\mathrm{m}$-sized and $1\;\mu\mathrm{m}$-sized monomers are $\simeq50\;\mathrm{m/s}$ and $\simeq7\;\mathrm{m/s}$, respectively. Collisions with mass ratios of 2-3 show two times smaller $\vfrag$ \citep[][]{Hasegawa2021}. Figure \ref{fig:summary_param} shows that coagulation instability efficiently concentrates dust grains and accelerates dust growth in both cases if $\alpha_0$ is on the order of $10^{-4}$ or smaller. The fragmentation velocity is lower for larger monomers. Coagulation instability will be inefficient in such a case. However, \citet{Tazaki2022} recently suggest that a monomer size is no greater than $0.4\;\mu\mathrm{m}$ based on optical and near-infrared polarimetric observations of several planet-forming disks. We thus expect that the above reference values are valid.

\citet[][]{Musiolik2016a,Musiolik2016b} show that the $\mathrm{CO}_2$ ice is ten times less sticky than $\mathrm{H}_2\mathrm{O}$ ice. This indicates $\vfrag\simeq5\;\mathrm{m/s}$ for $0.1\mu\mathrm{m}$-sized monomers if grains are fully covered by $\mathrm{CO}_2$ ice mantle \citep[see also the model adopted in][]{Okuzumi2019}. Even in such a case, coagulation instability can concentrate dust grains to promote dust growth according to Figure \ref{fig:summary_param}.

The necessary condition for efficient radial concentration of coagulation instability is weak turbulence of $\alpha_0<1\times10^{-3}$ as shown in Figure \ref{fig:summary_param}. Recently observed vertically thin dust disks might indicate such weak turbulence \citep[e.g.,][]{ALMA-Partnership2015,Andrews2018,Villenave2022}. \cite{Pinte2016} show $\alpha\simeq 3\times 10^{-4}$ for the HL Tau disk \citep{ALMA-Partnership2015}. Gas observations toward another disk (HD 163296) with $\mathrm{CO}$ and $\mathrm{DCO}^{+}$ lines also indicate weak turbulence of $\alpha\lesssim 3\times10^{-3}$ at high altitude and $\alpha\lesssim 1\times10^{-3}$ at low altitude \citep{Flaherty2015,Flaherty2017}. We can expect efficient coagulation instability in such weakly turbulent disks. Therefore, coagulation instability is a promising mechanism of icy planetesimal formation.

This work focuses on the dust disk evolution beyond the $\mathrm{H}_2\mathrm{O}$ snow line. Thus, we cannot rigorously discuss silicate-dust evolution via coagulation instability. Here, we just give brief comments focusing on the critical fragmentation velocity. According to \citet{Wada2009}, the critical fragmentation velocity of silicate dust aggregates is $\simeq 6\;\mathrm{m/s}$ for $0.1\mu\mathrm{m}$-sized monomers. Figure \ref{fig:summary_param} indicates that coagulation instability causes radial concentration and accelerates its growth even for such a relatively low $\vfrag$. Coagulation instability will efficiently concentrate silicate dust if the surface energy of silicate dust is ten times higher than previously assumed as indicated in \citet{Kimura2015} and \cite{Steinpilz2019}. Thus, the present results potentially indicate that coagulation instability also promotes rocky planetesimal formation at inner radii. Our future study will investigate nonlinear outcome of coagulation instability inside the $\mathrm{H}_2\mathrm{O}$ snow line.

\subsection{Possible evolution with self-induced dust trap}\label{subsec:selfinduced} 
 
 \citet{Gonzalez2017} found the self-induced dust trap whose development consists of the following two stages (see Section 4 therein). First, the drift deceleration due to the backreaction helps dust grains to grow larger, and the dust-rich region with $\taus\sim1$ forms. Second, the backreaction onto gas causes outward gas motion and forms a local pressure bump that traps dust grains further. Since we assume the steady gas disk, our simulations treat only the first stage of the self-induced dust trap. If we include the outward gas motion in the simulations, the self-induced dust trap will operate after the nonlinear coagulation instability, and dust grains are trapped more efficiently. This combined process promotes dust retention in a disk. Besides, coagulation instability sets up multiple locations where the self-induced dust trap operates while \citet{Gonzalez2017} observed the dust trap at a single location. Therefore, the combination of coagulation instability and the self-induced dust trap is a promising process to retain dust grains in a disk. In future work, we should further investigate whether or not the observationally estimated disk lifetime \citep[e.g.,][]{Strom1989,Ribas2014} can be explained in our scenario.
 
The outward gas motion due to the backreaction results in smaller pressure gradient (i.e., $\eta$), which reduces the dust drift. The reduction of $\eta$ around the dust ring can make the ring width larger \citep[][]{Kanagawa2018}. Thus, it may be expected that the combined process of coagulation instability and the self-induced dust trap creates wider rings than the rings forming only via coagulation instability. If rings become wide enough, those rings may be resolved by ALMA. Otherwise, higher-resolution observations are necessary to detect the rings forming via coagulation instability since the ring width is small ($\simeq0.5-1\;\mathrm{au}$, see Figure \ref{fig:ring_prop}).

\begin{figure*}[htp]
	\begin{center}
	\hspace{0pt}\raisebox{20pt}{
	\includegraphics[width=1.9\columnwidth]{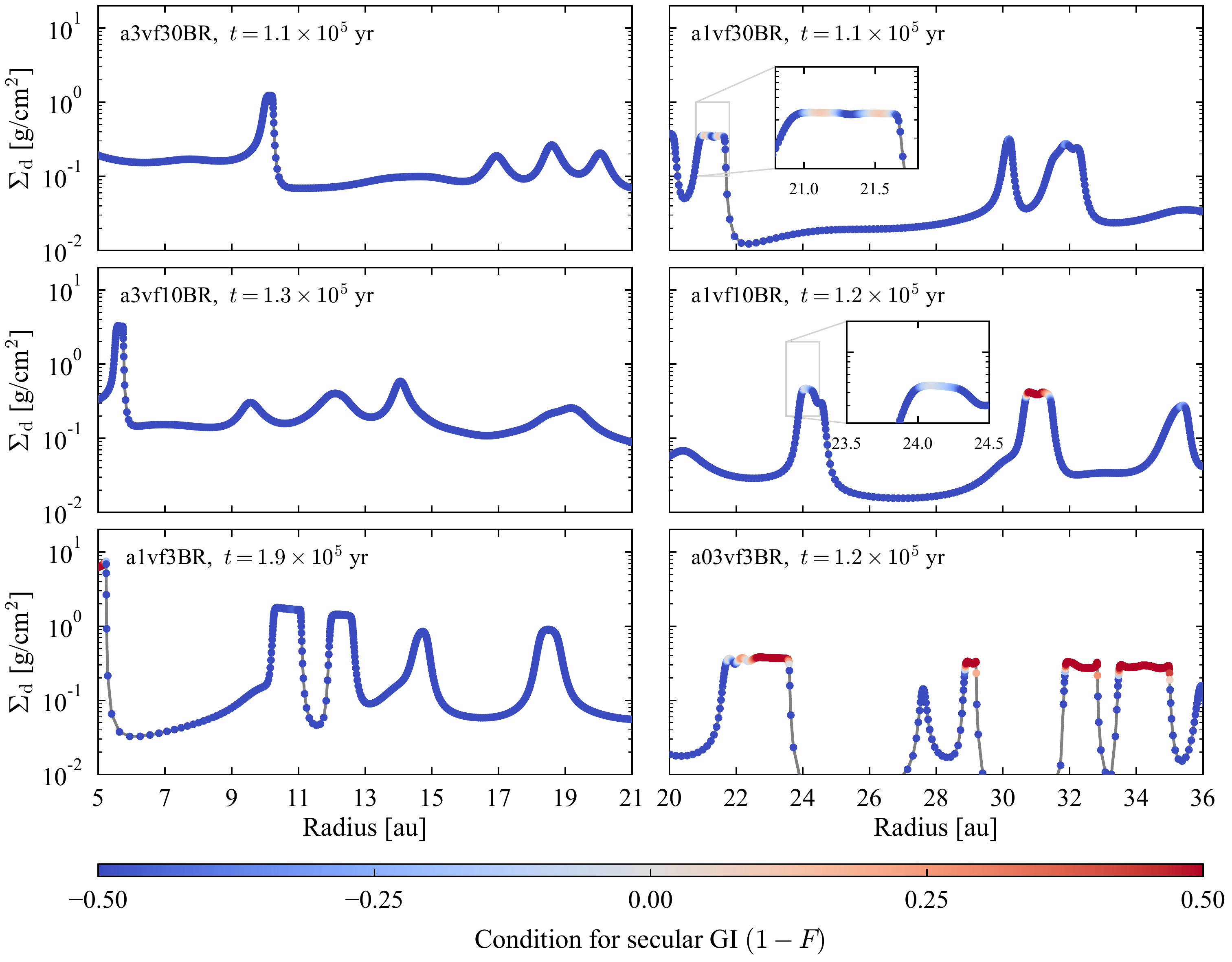}
	}
	\end{center}
	\vspace{-30pt}
\caption{Dust surface density profiles obtained in the six runs. The color of the markers represents whether secular GI is unstable or not at each radius. As found in Paper I, secular GI operates more easily in the outer region where Toomre's $Q$ is relatively low.}
\label{fig:SGI_cond}
\end{figure*}

\subsection{Implication for the pebble flux from outer regions}
Pebble accretion is one widely-investigated process for explaining the growth of a protoplanet in an inner region \citep[e.g.,][]{Ormel2010,Lambrechts2012,Lambrechts2014}\footnote{In this subsection, we use ``pebble" and ``dust" interchangeably although ``pebble" in this context should refer to solid particles of $\taus\sim 1$.}. It is argued that the pebble accretion is more efficient than the planetesimal accretion. However, it is also known that a massive pebble reservoir is prerequisite for a protoplanet to successfully grow larger with accreting pebbles. \citet{Lambrechts2019} show that the pebble reservoir of $110\merth$ is necessary to form terrestrial planets, and more massive reservoir of $340\merth$ is needed to form $10\merth$ planets \citep[see also][]{Bitsch2019}. Coagulation instability is expected to make the pebble accretion inefficient. The nonlinear development of coagulation instability causes the radial concentration at multiple radii and prevents pebbles from drifting inward. The pebble supply to the inner region will stop if the pebble trapping via coagulation instability efficiently occurs at outer radii (e.g., $> 10\;\mathrm{au}$). This thus potentially reduces the amount of pebbles available for the protoplanet growth in the inner region. In such a case, protoplanets will grow larger via planetesimal accretion \citep[e.g.,][]{Wetherill1989,Kokubo1998,Kobayashi2021}.

The present simulation is limited and only treats the formation of the first-generation rings before the drifting dust passes through the resulting ring. However, the efficiency of formation of second-generation rings and the radial migration of the remaining dust grains should affect the pebble flux. Longer-term simulations are thus necessary in future work to quantitatively examine the impact on the pebble accretion.

\subsection{Triggering other instabilities and planetesimal formation}\label{subsec:secular}
\citet{Tominaga2021} proposes a scenario that secular gravitational instability (GI) develops after coagulation instability toward planetesimal formation. We discuss whether or not secular GI is operational in the resulting dust rings in this subsection (see also Appendix C in Paper I). \citet{Tominaga2019} shows that the condition for the onset of secular GI is 
\begin{equation}
F\equiv \frac{Q^2\left(\taus\cd^2/\cs^2+\tilde{D}\right)}{(1+\varepsilon)\left[\taus\left(\varepsilon+\cd^2/\cs^2\right)+\tilde{D}(1+\varepsilon)\right]}<1,\label{eq:SGIcond1}
\end{equation}
where $\tilde{D}\equiv D\cs^{-2}\Omega$ and $\cd\propto\sqrt{\alpeff}\cs$ is velocity dispersion of dust grains \citep[e.g.,][]{YL2007}. The brief explanation is given in Appendix C in Paper I \citep[see also][]{Takahashi2014,Latter2017}. We calculate $1-F$ at each dust cell and investigate the stability of the resulting dust rings.

Figure \ref{fig:SGI_cond} shows the dust surface density profiles at the final time step from six runs. The color shows the values of $1-F$. The left panels show the results of the runs in which the densest rings form in relatively inner regions because of relatively strong diffusion (i.e. large $\alpha_0$). The right panels show the cases where $\alpha_0$ is relatively small and the densest rings form in relatively outer regions. We find that the outer rings tend to be unstable to secular GI. This trend is already discussed in Paper I: secular GI can grow more easily in outer regions where $Q$ is smaller (see Appendix C therein). From this point of view, weaker turbulence is preferable for secular GI to develop toward planetesimal formation after the dust concentration via coagulation instability. Because $F$ depends on the square of $Q$, a factor difference can change the stability of the rings, and thus massive gas disks are preferable. The decrease in $\alpeff$ at nonlinear growth phase also makes the rings more unstable to secular GI although the $\alpeff$-dependence of $F$ is weaker than $Q$-dependence. 

Using the approximate dispersion relation of secular GI derived by \citet{Tominaga2019} (see Equations (26)-(31) therein), we find that the maximum growth rate of secular GI, $n_{\mathrm{SGI,max}}$, is roughly given by
\begin{align}
\frac{n_{\mathrm{SGI,max}}}{\Omega}&\sim\frac{1-F}{F}\frac{1+\varepsilon}{1+\varepsilon\cd^2/\cs^2}\frac{\taus\left(\varepsilon+\cd^2/\cs^2\right)+(1+\varepsilon)\tilde{D}}{\taus^2+(1+\varepsilon)^2}\notag\\
&\sim\frac{1-F}{F}\frac{\taus\varepsilon}{\taus^2+1},
\end{align}
where we assume $\varepsilon\ll 1$, $\tilde{D}\sim\cd^2/\cs^2\ll\taus\varepsilon$, and we estimate the most unstable wavenumber $k$ as $kH\simeq QF^{-1}/(1+\varepsilon)$\footnote{ \citet{Tominaga2019} showed that the approximated dispersion relation well reproduces the exact dispersion relation for $\varepsilon=0.1$ (see Figure 2 therein). Since the resulting rings have $\varepsilon\sim0.01$, the use of the approximated equation is valid. }. The maximum growth rate can be further reduced to $n_{\mathrm{SGI,max}}/\Omega\sim 0.5\varepsilon(1-F)F^{-1}$ since dust grains grow to the size of $\taus\sim1$ via coagulation instability. For $F=0.5$ and $\varepsilon=0.01$, secular GI develops within a few tens Keplerian periods. This timescale is comparable to the dust growth timescale for $\varepsilon=0.01$ \citep[e.g., see Equation (38) in][]{Brauer2008}. Besides, the growth timescale of secular GI can be shorter at the ring than the ring-drift timescale $r/|v_r|$ (e.g. $\sim3\times10^{4}\;\mathrm{yr}$ for the ring at 31 au in the a1vf10BR run. see Section \ref{sec:results}). Therefore, secular GI in the resulting rings certainly affects the dust-disk evolution and promotes planetesimal formation. 

We note that there are rings in the middle of growing at the final time step (e.g., the ring at $r\simeq24\;\mathrm{au}$ in the a1vf10BR run). Those rings will become unstable to secular GI once the dust concentration and the dust growth proceed enough. The development of secular GI will lead to planetesimal formation through further radial dust concentration \citep[][]{Tominaga2020} and azimuthal fragmentation \citep[][]{Pierens2021}. Therefore, a combination of coagulation instability and secular GI is one promising mechanism for planetesimal formation.

\citet{Tominaga2019} find another secular instability called two-component viscous GI (TVGI). TVGI operates more easily than secular GI \cite[e.g., see Figure 8 in][]{Tominaga2019} if the dust drift is insignificant \citep[][]{Tominaga2020}. As mentioned in Section \ref{subsec:selfinduced}, the pressure gradient will be reduced as a result of radial concentration via coagulation instability and the self-induced dust trap. This indicates that TVGI can be also operational in the resulting rings, and planetesimal formation is possible in a wider parameter space. We will address this process in our future studies.

Strong dust clumping by streaming instability is also regarded as a promising mechanism of planetesimal formation \citep[e.g.,][]{YoudinGoodman2005,Youdin2007,Johansen2007,Johansen2007nature}. The spatial scale of streaming instability is much smaller than the spatial scale of secular GI, and thus both instabilities can operate at the same time\footnote{Turbulent motion due to streaming instability is implicitly assumed in this work through the $\epmid$-dependence of $\alpha$.}. The previous studies investigated the required dust-to-gas ratio for the strong clumping to operate via streaming instability \citep[][]{Carrera2015,Yang2017,Li2021}. The often-used approximated criteria is $\epmid\gtrsim1$ although \citet{Li2021} found that the critical $\epmid$ is smaller ($\simeq0.3-0.8$) for $0.02\lesssim\taus\lesssim 1$ (see Figure 4 therein). Rings forming via coagulation instability are preferable locations for streaming instability to grow efficiently since $\epmid$ becomes $\sim1$ . From this point of view, we may expect the onset of streaming instability in the rings. On the other hand, \citet{Carrera2015} found that the strong clumping via streaming instability requires larger $\sigmad/\sigmag$ for $\taus\geq1$. This may indicate that streaming instability in the rings becomes inefficient since $\taus$ increases toward unity via nonlinear coagulation instability and will become larger.

 We should also note that recent studies showed dependence of the efficiency of streaming instability on the dust size distribution \citep[e.g.,][]{Krapp2019,Paardekooper2020,Paardekooper2021,McNally2021,Zhu2021,Yang2021}. \citet{Krapp2019} showed that the linear growth rate of streaming instability can be much smaller when one assumes a power-law size distribution. \citet{McNally2021} showed that the linear growth rate is less reduced for a power-law-bump distribution where they assume a Gaussian bump at the top end of a power-law distribution (see Figure 8 therein). This indicates that the dust segregation before the onset of streaming instability is important to discuss the possible connection from coagulation instability. In the present simulations, we only treat the representative dust size with the assumption of collisions of the size ratio of 0.5 \citep[][]{Sato2016}. Possible dust segregation during the nonlinear development of coagulation instability and a combined process with streaming instability should be studied in more detail, which is beyond the scope of this paper.

After dust grains grow enough, the rings can be unstable to the classical GI \citep[e.g.,][]{Goldreich1973,Sekiya1983}. \citet{Michikoshi2016,Michikoshi2017} calculated the random velocity of icy dust aggregates as a function of $\rhoint$ and $\mpar$ and investigated the stability of a dust layer \citep[see][for silicate-dust cases]{Tatsuuma2018}. The aggregate mass range considered in their studies is $\mpar\geq10^8\;\mathrm{g}$, corresponding to $\gtrsim10^{2.5}\;\mathrm{cm}$ for compact dust with $\rhoint\sim1\;\mathrm{g/cm}^3$. In our disk model, dust grains in this size range have $\taus>1$, which will form after nonlinear coagulation instability. Their results indicate that $\alpha\lesssim10^{-3}$ is sufficient for dust GI to take place during the dust-size evolution for $\mpar\geq10^{8}\;\mathrm{g}$ \citep[e.g., see Figure 6 in][]{Michikoshi2017}. Therefore, the dust rings observed in the present simulations ($\Sigma_{\gas,10}=2\sigmagmmsn(10\;\mathrm{au}$)) will become unstable to the dust GI in the sense of the Toomre criterion after dust becomes large enough ($\taus>1$). In this case, secular GI and streaming instability may not be required for planetesimal formation. Gradual dust growth in the resulting rings will naturally cause the classical GI and formation of planetesimals.

\section{Summary}\label{sec:summary}

Planetesimal formation is the first step in the planet forming process. However, it is known that the radial drift and the collisional fragmentation can limit dust growth and prevent the formation of planetesimals. One promising mechanism is hydrodynamical dust clumping due to disk instabilities (e.g., streaming instability and secular GI) and subsequent planetesimal formation via GI of the resulting dust clouds or dust rings. Our previous study based on a linear analysis proposes coagulation instability as a promising mechanism for the dust clumping \citep{Tominaga2021}. The series of the studies in Paper I and Paper II investigates the nonlinear outcome of coagulation instability. This paper (Paper II) focuses on the nonlinear development and the dust ring formation via coagulation instability under the influence of the backreaction (the drift deceleration) and the fragmentation. The results and the findings are listed below:
\begin{itemize}
\item In  the absence of the backreaction, fragmentation limits dust growth in the rings ($\taus<1$) and saturates the nonlinear growth of the instability, and as a result the rings suffer the fast drift.
\item Previous studies found that the backreaction weakens turbulence and enhances the dust settling \citep[e.g.,][]{Takeuchi2012,Lin2019,Xu2022}. The enhanced settling increases a local dust-to-gas ratio and further augments the backreaction, which lead to a positive feedback as indicated in \citet[][]{Takeuchi2012}. We adopt a simple model of $\epmid$-dependent turbulence strength and investigate a combined process with coagulation instability. In the presence of the backreaction, the drift speed is reduced as dust grains concentrate, and the resulting rings have a plateau and relatively wide structures in the dust surface density profile (e.g., Figures \ref{fig:evol_fid} and \ref{fig:ring_prop}) especially for small $\alpha_0$.
\item The combination of the dust concentration via coagulation instability and the backreaction enables dust growth toward the size of $\taus=1$ (see also Figures \ref{fig:collvel} and \ref{fig:collvel_a3vf10BR}). Dust grains will be retained because of the reduced drift velocity due to the backreaction. We should further investigate whether or not the observationally estimated disk lifetime can be explained in the present scenario.
\item The ring mass of $\simeq0.5\merth-1.5\merth$ is the most frequent in the present runs with $\Sigma_{\gas,10}=2\sigmagmmsn(10\;\mathrm{au})$ (Figures \ref{fig:ringmass_a1vf10BR} and \ref{fig:mass_cdf_param}). The ring mass will be smaller for a less massive disk with a fixed initial dust-to-gas surface density ratio since the most unstable wavelength depends $\sigmad$ only through the surface density ratio \citep[][]{Tominaga2021}.
\item The necessary condition for the radial dust concentration and the accelerated dust growth is $\alpha< 1\times10^{-3}$ and $\vfrag> 1\;\mathrm{m/s}$ (Figure \ref{fig:summary_param}).
\end{itemize}

According to the previous studies on the sticking properties of water ice, the critical fragmentation velocity is $\simeq50$ m/s and $\simeq7$ m/s for monomer sizes of 0.1 and 1 $\mu$m, respectively \citep[e.g.,][]{Wada2009}. We can expect the ring formation and the accelerated dust growth via coagulation instability in such cases if $\alpha_0$ is on the order of $10^{-4}$ or less. Recently, CO$_2$ ice is found to be less sticky than water ice \citep[e.g.,][]{Musiolik2016a,Musiolik2016b}, and the critical velocity is less than 10 m/s \citep[][]{Pinilla2017,Okuzumi2019} if dust is fully covered by CO$_2$ ice mantle \citep[cf.][]{Kouchi2021}. In such a case, the turbulence strength of $\alpha<3\times10^{-4}$ is required for coagulation instability to develop. 

We find that the dust-to-gas ratios at the midplane and in the dust sublayer are on the order of unity in most cases in the ring (e.g., see the right panel of Figure \ref{fig:maximumDG} for $\epmid$). Although our simulations assume steady gas profile, the backreaction will modify the gas profile around the dust rings, which triggers self-induced dust trap \citep[][]{Gonzalez2017} at multiple radii. We expect that a combination of coagulation instability and the subsequent self-induced dust trap is a promising process to retain dust grains (Section \ref{subsec:selfinduced}). Such dust trapping at multiple radii will reduce the inward pebble flux, indicating that the so-called pebble accretion might become inefficient if coagulation instability efficiently develops.

The resulting rings can be unstable to other instabilities. We find that secular GI can operate in the rings in the outer region where Toomre's $Q$ for gas is relatively small (Figure \ref{fig:SGI_cond}). Even when secular GI is stable in the ring, gradual dust size growth ($\mpar\geq 10^8\;\mathrm{g}$) can trigger the classical GI \citep[][]{Michikoshi2016,Michikoshi2017}. These combined processes with other mechanisms further promote dust clumping and dust growth, and will finally lead to planetesimal formation. Therefore, we expect that the radial dust concentration and the enhanced dust growth via coagulation instability is the key process for planetesimal formation. In particular, coagulation instability has the potential to trigger outer planetesimal formation at $r\geq10\;\mathrm{au}$, where pure dust growth is inefficient.

\acknowledgments
We thank Elijah Mullens for fruitful discussion and helpful comments. We also thank the anonymous referee for carefully reading the manuscript and providing constructive comments, which helped us to improve the manuscript. This work was supported by JSPS KAKENHI Grant Nos. JP18J20360, 21K20385 (R.T.T.), 19K03941 (H.T.), 16H02160, 18H05436, 18H05437 (S.I.), 17H01103, 17K05632, 17H01105, 18H05438, 18H05436 and 20H04612 (H.K.). R.T.T. is also supported by RIKEN Special Postdoctoral Researchers Program.

%




\appendix
\section{Timescale of Radial ``Dispersal" of A Ring}\label{app:disp}
In the present study, utilizing the moment approach \citep{Sato2016}, we focus on the radial motion and the size growth of the peak mass $\mpar$ for simplicity. In this appendix, we estimate a timescale of ring evolution by roughly taking the effect of size dispersion into account and discuss the impact on the dust evolution after the ring formation via coagulation instability.

For simplicity, we assume two-population dust grains with $\taus=\tau_{\mathrm{s},1}$ and $\tau_{\mathrm{s},2}$. We regard the dust of $\taus=\tau_{\mathrm{s},1}$ as the peak-mass dust. As noted in \citet{Sato2016}, the peak mass evolution is well reproduced if we consider collisions with the size ratio of 0.5. We thus assume $\tau_{\mathrm{s},2}=0.5\tau_{\mathrm{s},1}$, and call this population ``small dust". The total dust surface density $\sigmad$ is given by 
\begin{equation}
\sigmad = \Delta\sigmad(\tau_{\mathrm{s},1}) + \Delta\sigmad(0.5\tau_{\mathrm{s},1})
\end{equation}
where $\Delta\sigmad(\tau)$ is the surface density of dust with $\taus=\tau$. 

We consider time evolution of a ring consisting of the peak-mass dust and the small dust. These two dust grains have different radial drift velocity, and thus the small dust flows out of the ring in the rest frame of the peak-mass dust. This ``dispersal" decreases $\sigmad$ of the ring, which we call ``ring dispersal". The time evolution of $\sigmad$ is roughly given by
\begin{equation}
\left(\frac{\partial\sigmad}{\partial t}\right)_{\mathrm{disp}} = -\Delta\sigmad(0.5\tau_{\mathrm{s},1})\frac{|v_r(\tau_{\mathrm{s},1})-v_r(0.5\tau_{\mathrm{s},1})|}{\Delta R},
\end{equation}
where $\Delta R$ is the ring width and is $\simeq0.5\;\mathrm{au}-1\;\mathrm{au}$ in our simulations (e.g., see Figure \ref{fig:ring_prop}). Thus, the typical timescale of the ring dispersal $t_{\mathrm{disp}}$ is
\begin{align}
t_{\mathrm{disp}}&\equiv\sigmad\left(\frac{\partial\sigmad}{\partial t}\right)_{\mathrm{disp}}^{-1}\notag\\
&=\frac{2\Delta R}{|v_r(\tau_{\mathrm{s},1})|}\left(\frac{\sigmad/\Delta\sigmad(0.5\tau_{\mathrm{s},1})}{2}\right)\left(\frac{|v_r(\tau_{\mathrm{s},1})-v_r(0.5\tau_{\mathrm{s},1})|}{|v_r(\tau_{\mathrm{s},1})|}\right)^{-1}.
\end{align}
From the definition of the peak mass, $\Delta\sigmad(0.5\tau_{\mathrm{s},1})$ is smaller than $0.5\sigmad$. We may expect that the last term is on the order of $0.5$. We then obtain $t_{\mathrm{disp}}\gtrsim 4\Delta R/|v_r(\tau_{\mathrm{s},1})|$. 

\begin{figure}[htp]
	\begin{center}
	\hspace{0pt}\raisebox{20pt}{
	\includegraphics[width=0.4\columnwidth]{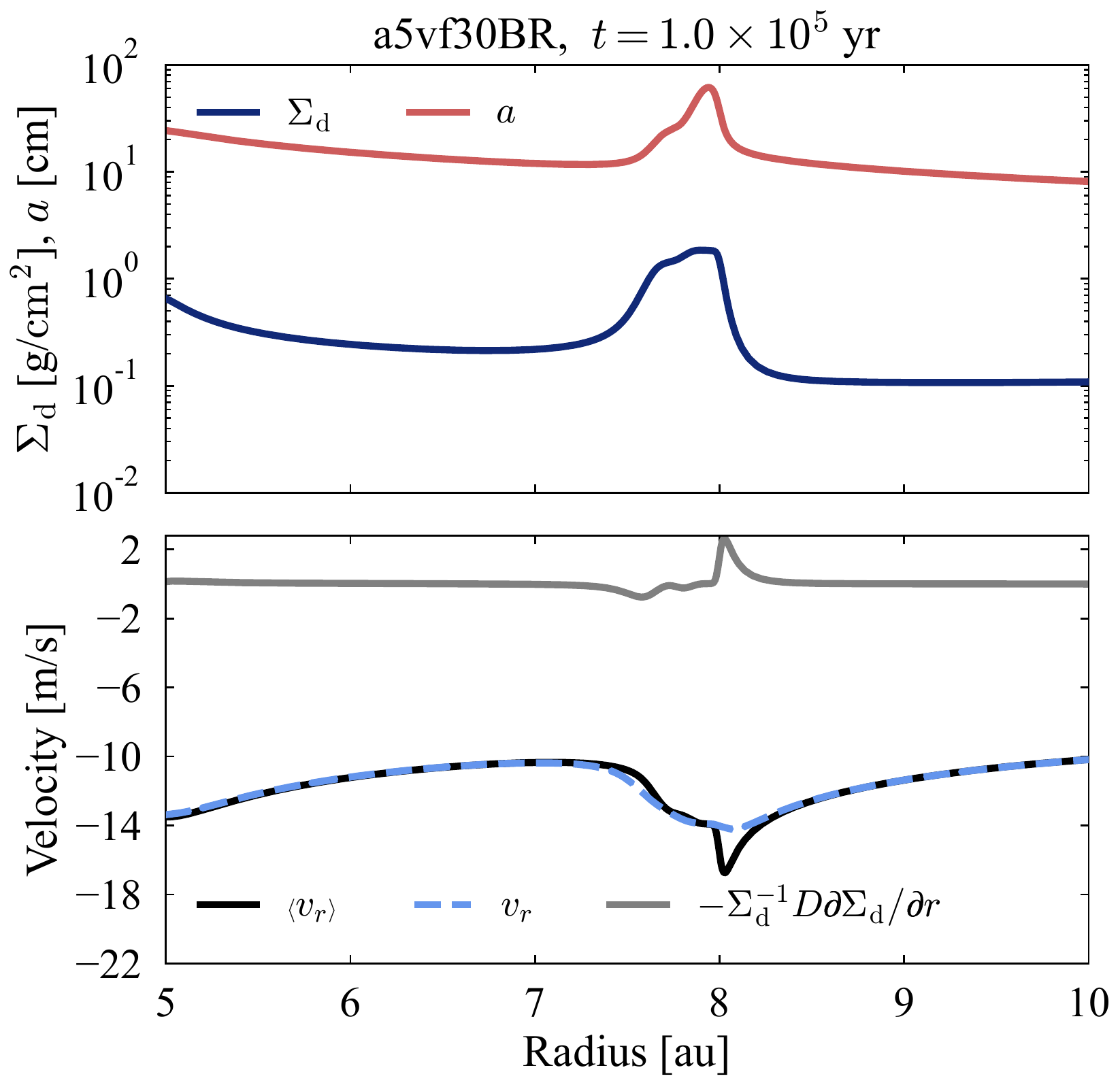}
	}
	\end{center}
	\vspace{-30pt}
\caption{The resulting ring structure in a5vf30BR run. (Top panel) The radial profiles of dust surface density (dark blue line) and dust size $a$ (red line) are shown. (Bottom panel) The radial profiles of the mean drift velocity $\left<v_r\right>$ (black line), the total velocity $v_r$ (blue dashed line), and the diffusion velocity (gray line) are shown.}
\label{fig:ring_prop_a5vf30}
\end{figure}

In the case of a1vf10BR run, the ring width $\Delta R$ is $\simeq1\;\mathrm{au}$, and the radial velocity $|v_r(\tau_{\mathrm{s},1})|$ is $\simeq1\times10^{-3}\;\mathrm{au/yr}$. We then obtain $t_{\mathrm{disp}}\gtrsim 4\times 10^{3}\;\mathrm{yr}$, which is longer than coagulation timescale at 31 au. In this case, we may expect that the size dispersion insignificantly affect the dust-ring evolution. The timescale $\Delta R/|v_r(\tau_{\mathrm{s},1})|$ can be smaller for large $\vfrag$ and large $\alpha$ (see Figure \ref{fig:maximumDG}). For example, we find in a5vf30BR run that the velocity of a ring forming at 8 au is $\simeq14\;\mathrm{m/s}\simeq3\times10^{-3}\;\mathrm{au/yr}$, and the ring width is $\Delta R\simeq 0.5\;\mathrm{au}$ (Figure \ref{fig:ring_prop_a5vf30}). Adopting these values, we obtain $t_{\mathrm{disp}}\gtrsim 7\times 10^{2}\;\mathrm{yr}$. Nevertheless, the coagulation timescale at the ring position ($r\simeq8\;\mathrm{au}$) is $\simeq3\times10^2\;\mathrm{yr}$ and is still shorter than the dispersal time. Therefore, we expect that the dust growth in the ring proceeds enough before the ring is dispered. More detailed investigation with the dust size distribution is our future study.

\section{Dependence of the sublayer thickness}\label{app:zsdepend}
In the present simulations, we assume that the momentum transfer between dust and gas takes place within the sublayer. Its vertical extent is assumed to be $|z|\leq\hd$. The vertical extent would depend on detailed turbulence structure that is not treated in the present 1D simulation. In this appendix, we give brief comments on the dependence on the sublayer thickness.

\begin{figure*}[htp]
	\begin{center}
	\hspace{0pt}\raisebox{20pt}{
	\includegraphics[width=0.8\columnwidth]{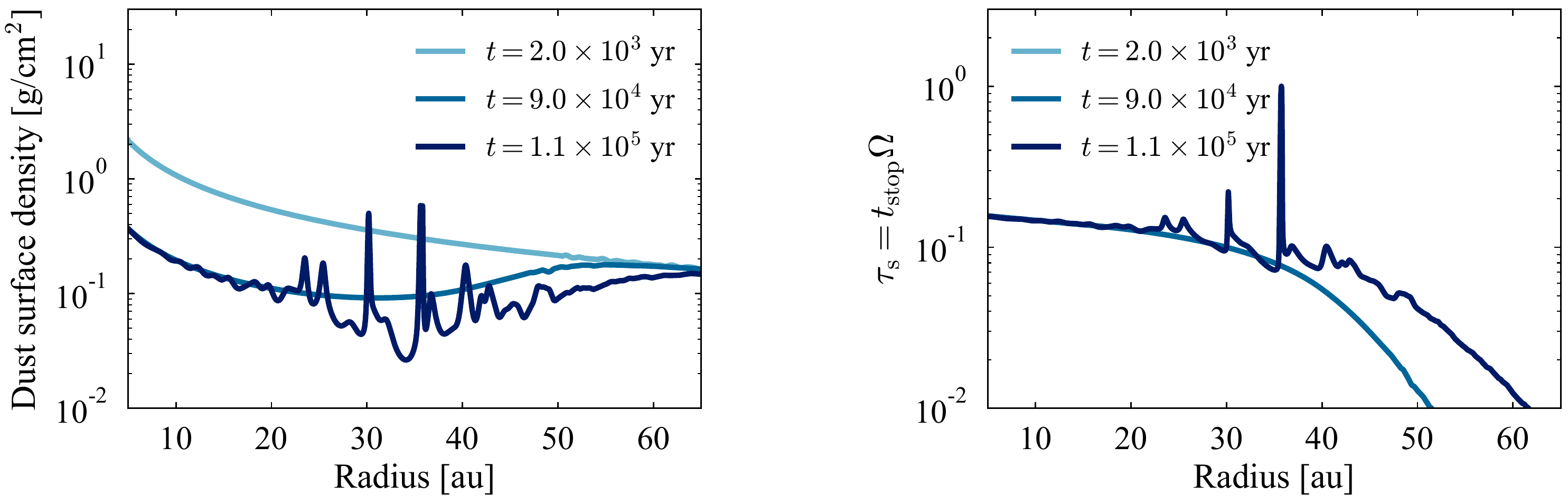}
	}
	\end{center}
	\vspace{-30pt}
\caption{The results of a1vf10BR run with the sublayer thickness of $|z|\leq 3\hd$. The left and right panels show the time evolution of $\sigmad$ and $\taus$. We note that $\taus$ is less than $10^{-2}$ at $t=2\times 10^3\;\mathrm{yr}$, and thus the light blue line does not appear on the right panel. The dust-to-gas ratio in the ring is larger than the run with the sublayer thickness of $|z|\leq\hd$ (Figure \ref{fig:evol_fid}). This is because the drift deceleration due to the backreaction is weaker and coagulation instability grows more efficiently. }
\label{fig:a1vf10BR_z3}
\end{figure*}

We conduct a run with the same parameters and initial perturbations as in a1vf10BR run but with the sublayer extent of $|z|\leq3\hd$. Figure \ref{fig:a1vf10BR_z3} shows the results. We find that coagulation instability grows faster and the dust-to-gas ratio in the densest ring becomes larger. We attribute this to the reduction of the drift deceleration due to the backreaction. For the larger vertical extent, the dust-to-gas ratio in the sublayer $\epsub$ becomes smaller for a given surface density ratio $\varepsilon$ (Equation (\ref{eq:epsub})), and thus the drift velocity is closer to the test-particle limit (see Equation (\ref{eq:vrdrift})). The efficiency of coagulation instability is higher for weaker backreaction \citep[see also Section 4.2 in][]{Tominaga2021}. This trend is consistent with the $\alpha_0$-dependence of the resulting $\varepsilon$ (Figure \ref{fig:maximumDG}): larger $\alpha_0$ stirs dust grains up more efficiently, reduces the drift deceleration, and leads to larger $\varepsilon$. Thus, we expect efficient coagulation instability when the momentum transfer between dust and gas takes place within a relatively large vertical extent (e.g., a few dust scale heights). We defer more detailed discussion to our future study.

\section{On the ring formation near the inner boundary}\label{app:on_a3vf10br}
\begin{figure*}[htp]
	\begin{center}
	\hspace{0pt}\raisebox{20pt}{
	\includegraphics[width=0.5\columnwidth]{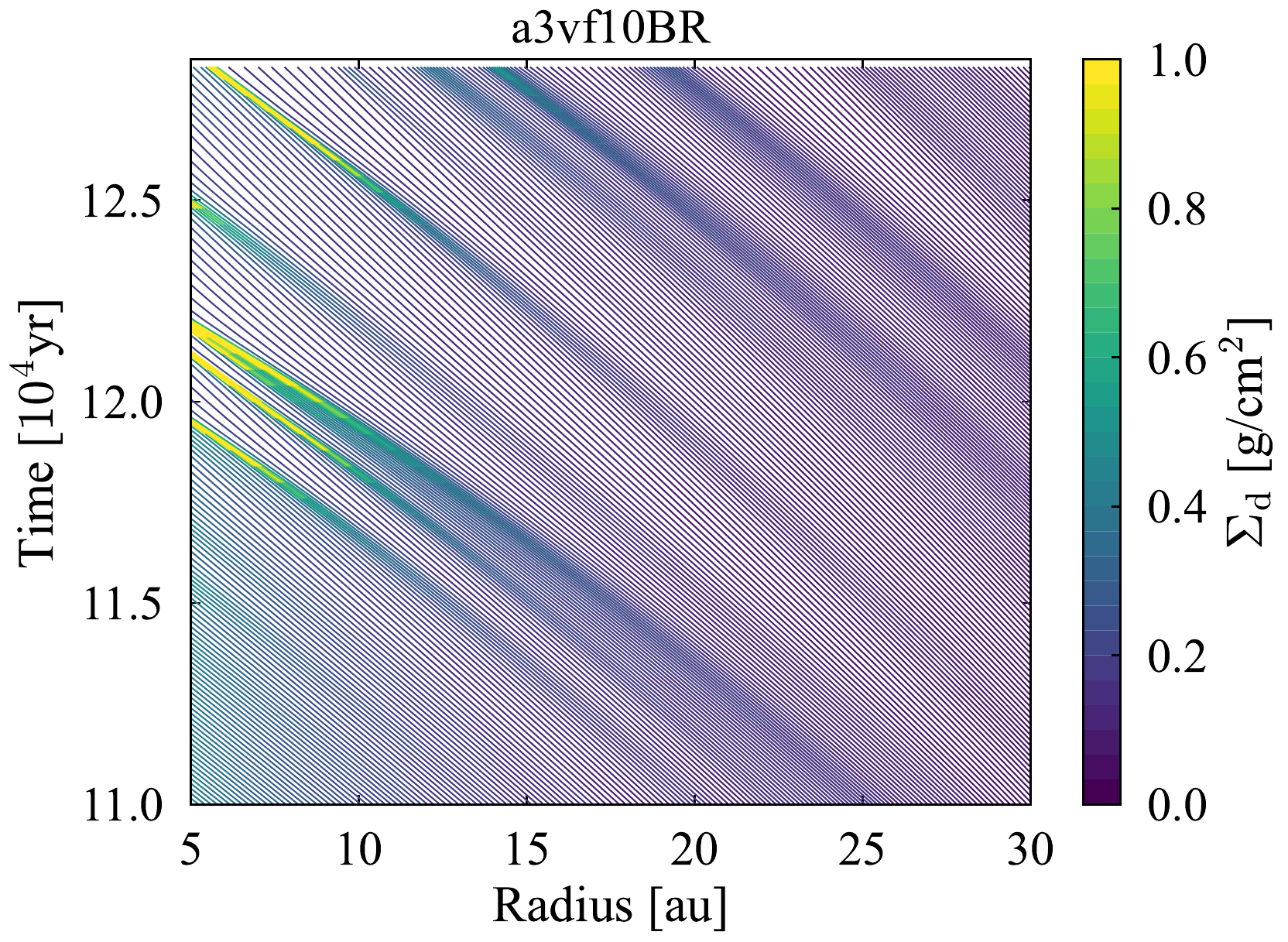}
	}
	\end{center}
	\vspace{-30pt}
\caption{Trajectories of dust cells seen in the a3vf10BR run. Color represents the dust surface density at each dust cell. As mentioned in Figure \ref{fig:dvtot_on_r-t}, the dust surface density is larger at a radius where more dust cells gather since our numerical method is based on the Lagrangian cells. There are multiple rings, and the dust surface density of the ring monotonically increases as the ring moves inward. It is important to note that the rings start growing in outer regions that are far from the inner boundary. Therefore, the ring formation near the inner boundary in Figure \ref{fig:a3vf10BR} is not due to a boundary effect.  }
\label{fig:a3vf10BR_cell_tracking}
\end{figure*}

 Figure \ref{fig:a3vf10BR} shows that the ring forms near the inner boundary ($r=5\;\mathrm{au}$). To show that this is not due to a boundary effect, we plot trajectories of dust cells of the a3vf10BR run in Figure \ref{fig:a3vf10BR_cell_tracking}. Color represents the dust surface density. One can see that multiple perturbations radially drift from outer radii and some of them reach the inner boundary until the final timestep ($t\simeq1.28\times10^5$ yr). The most developed ring in Figure \ref{fig:a3vf10BR} is located at $r\simeq30\;\mathrm{au}$ for $t\simeq1.12\times 10^5$ yr, which is far from the inner boundary. The surface density of the ring monotonically increases as the ring moves inward. In this way, the observed ring formation near the inner boundary is not due to a boundary effect.

\bibliographystyle{aasjournal}
\bibliography{rttominaga2021c}

%
%
%


\end{document}